\documentclass[authoryear,preprint,review,12pt]{elsarticle}

\usepackage{amsthm}
\usepackage{amsmath}
\usepackage{amsfonts}
\usepackage{amssymb}
\usepackage{graphicx}
\usepackage{enumerate}
\usepackage{color}
\usepackage{array}
\usepackage{bbm,bm}
\usepackage{multirow}
\usepackage{rotating}
\usepackage{xr}
\usepackage{setspace}
\usepackage{pdflscape}
\usepackage{afterpage}

\usepackage{url} % not crucial - just used below for the URL 

\usepackage{hyperref}
\usepackage{xcolor}
\definecolor{Red}{rgb}{0.5,0,0}
\definecolor{Blue}{rgb}{0,0,0.5}
\definecolor{darkgreen}{rgb}{0.0, 0.5, 0.0}  % Dark green
\hypersetup{%
	colorlinks = {true},
	linktocpage = {true},
	plainpages = {false},
	linkcolor = {Blue},
	citecolor = {Blue},
	urlcolor = {Red},
	pdfstartview = {Fit},
	pdfpagemode = {UseOutlines},
	pdfview = {XYZ null null null}
}

\begin{document}

\begin{frontmatter}



%% Article title
\title{Joint modeling of low and high extremes using a multivariate extended generalized Pareto distribution} 

% ---- Authors 
\author[aff1,aff2]{Noura Alotaibi}
\ead{noura.alotaibi@kaust.edu.sa} 
\author[aff1]{Matthew Sainsbury-Dale} \ead{matthew.sainsburydale@kaust.edu.sa} 
\author[aff3]{Philippe Naveau} \ead{philippe.naveau@lsce.ipsl.fr} 
\author[aff4]{Carlo Gaetan} 
\ead{gaetan@unive.it} 
\author[aff1]{Rapha\"el Huser\corref{cor1}} \ead{raphael.huser@kaust.edu.sa} 

\cortext[cor1]{Corresponding author.} 

\address[aff1]{Statistics Program, CEMSE Division, King Abdullah University of Science and Technology (KAUST), Thuwal 23955--6900, Saudi Arabia} 
\address[aff2]{Department of Mathematics, College of Science, Imam Abdulrahman Bin Faisal University, Dammam 31113, Saudi Arabia} 
\address[aff3]{Laboratoire des Sciences du Climat et de l’Environnement, Bat 714 -- ICE, 91191 Gif-sur-Yvette, France} 
\address[aff4]{Department of Environmental Sciences, Informatics and Statistics, Ca' Foscari University of Venice, 30123 Venice, Italy}


%\author{} %% Author name

%\affiliation{organization={},%Department and Organization

%% Abstract
\begin{abstract}
In most risk assessment studies, it is important to accurately capture the entire distribution of the multivariate random vector of interest from low to high values. For example, in climate sciences, low precipitation events may lead to droughts, while heavy rainfall may generate large floods, and both of these extreme scenarios can have major impacts on the safety of people and infrastructure, as well as agricultural or other economic sectors. In the univariate case, the extended generalized Pareto distribution (eGPD) was specifically developed to accurately model low, moderate, and high precipitation intensities, while bypassing the threshold selection procedure usually conducted in extreme-value analyses. In this work, we extend this approach to the multivariate case. The proposed multivariate eGPD has the following appealing properties: (1) its marginal distributions behave like univariate eGPDs; (2) its lower and upper joint tails comply with multivariate extreme-value theory, with key parameters separately controlling dependence in each joint tail; and (3) the model allows for fast simulation and is thus amenable to simulation-based inference. We propose estimating model parameters by leveraging modern neural approaches, where a neural network, once trained, can provide point estimates, credible intervals, or full posterior approximations in a fraction of a second. Our new methodology is illustrated by application to daily rainfall times series data from the Netherlands. The proposed model is shown to provide satisfactory marginal and dependence fits from low to high quantiles.
\end{abstract}



%% Keywords

\end{frontmatter}




\allowdisplaybreaks

%%%%%%%%%%%%%%%%%%%%%%%%%%%%%%%%%
%%%%%%%%%%%%%%%%%%%%%%%%%%%%%%%%%
\section{Introduction}

Record-breaking extreme weather events stemming from opposite distributional tails, such as heat and cold waves or floods and droughts, can both lead to major impacts and pose serious health, social, environmental, and economic risks \citep{ipcc2023,fischer2025}. 
%For example, heatwaves are known to pose serious health, social, environmental, and economic risks \citep{Zhong.etal:2021,Thompson.etal:2023}. While multiple definitions of a heatwave have been proposed in the literature, the World Meteorological Organization (see \url{https://wmo.int/topics/heatwave}) defines it as a period of abnormally hot weather where both the maximum and minimum temperatures at a given location are unusually high for a few consecutive days. On the opposite side of the spectrum, coldwaves (such as the 2022 North American winter storm, see \citealp{Gong.etal:2024}) are defined as periods of extraordinarily low temperatures and can similarly cause fatalities, affect the survival of certain animal species \citep{thibaud2016Bayesian}, and cause major disruptions, e.g., through the increased pressure on the energy production sector \citep{Juan.etal:2017}. The relevance of studying extreme events from both tails is not limited to temperature data. 
%Another important climate variable is precipitation, excess of which may lead to flooding \citep{Brunner.etal:2020,Merz.etal:2021,Jun.etal:2022}, whereas lack of which may lead to droughts \citep{Vicente-Serrano.etal:2020}. 
Widespread flooding, for example, affects people worldwide \citep{Brunner.etal:2020} and is one of the costliest hazards in the United States according to \citet{SwissRe:2019}. Among other geo-environmental factors, flooding often results from excess precipitation, which can either be caused by a single heavy rainfall event or by several consecutive but potentially less extreme rainfall events, whose accumulation may saturate the absorption ability of the ground \citep{Merz.etal:2021,Jun.etal:2022}. On the opposite side of the spectrum, droughts often cause major economic disruptions through their impact on agriculture and can also have health impacts due to decreased water accessibility (especially in low-income countries), among other factors. Droughts are usually defined as a period of time during which there is a lack of, or abnormally low levels of, precipitation combined with extreme heat \citep{Vicente-Serrano.etal:2020}. Therefore, to study the potential impacts of precipitation-related extreme events, it is important to build statistical models that can accurately capture the behavior of precipitation intensities all the way from low values in the lower tail to high values in the upper tail, while also flexibly describing behavior in the bulk of the distribution. % \citep{naveau2016modeling}.

From a statistical perspective, extreme events may be studied through the lens of extreme-value theory which characterizes the possible tail behaviors under certain regularity conditions \cite[see, e.g., the books of ][]{coles_introduction_2001,beirlant:goegebeur:teugels:segers:2004,embrechts2013modelling,Carvalho25}. 
In the univariate framework, the generalized Pareto distribution plays a central role, being the only possible upper-tail limit of (properly rescaled) high threshold exceedances as the threshold increases arbitrarily \citep{Davison.Smith:1990}. An analogous representation can be obtained for the lower tail by symmetry, describing the behavior of extremely low threshold non-exceedances. When interest lies in both tails, a classical approach has thus been to fit the generalized Pareto distribution to extreme data from each tail separately; the fitted generalized Pareto distributions are then used for risk assessment by extrapolating further into the tails. While this approach has the benefit of being simple (under the assumption that appropriately chosen generalized Pareto distributed exceedances have been selected), it is based on the premise that the bulk of the distribution does not play an important role and that its analysis can be decoupled from that of (moderate) extremes. In practice, however, the bulk of the distribution often provides crucial information.
% For example, 
% stochastic weather generators (i.e., emulators) of the entire rainfall range \cite[see, e.g.,][]{Obakrim25} are used to produce ensembles of possible time series that serve as inputs to assess the sensitivity of crop computer models. 
% The outputs of such crop models are influenced by the distributional features of low, moderate and heavy rainfall.
Moreover, when two generalized Pareto distributions and another distribution are fitted separately to low extremes, high extremes and the bulk, respectively,  
%another distribution is fitted to data from the bulk in order to complete the picture, 
a major drawback is to determine how to put these three pieces back together. 
The resulting bulk-tails mixture model will be discontinuous at the two chosen thresholds, which is an unwelcome artificial feature not present in the original data.  In addition, threshold selection to define extreme events has always been a delicate issue in extreme-value theory  \cite[see, for example,  the so-called ``horror Hill plots'' discussed in][]{ResnickStarica97}. % not trivial as it determines where and how the bulk transitions into  tails, and 
Assessing, both practically and theoretically, how threshold selection impacts downstream statistical inference is difficult \citep{Scarrott.MacDonald:2012,Alouini2026}. 
Sensitivity with respect to the threshold choice can be visually explored in the univariate context, but such ad-hoc visual devices (like mean-excess function and quantile-quantile plots) can become impractical in a multivariate context. In practice, the same quantile---e.g., the 95\% percentile---is often arbitrarily chosen for all marginal variables \cite[see, e.g.,][]{Davison.Huser:2015}. 

To mitigate these practical issues, various modeling approaches have been proposed in the univariate case \cite[see][for a recent review]{Naveau25}. 
%- this can be solved by imposing continuity (e.g., Gaussian+GPD) but the model becomes more constrained and thus less flexible (Carreau and Bengio).\\
For positive random variables with a heavy upper tail, a first model proposed by 
\cite{Frigessi2002} 
was constructed as a mixture between a light-tailed distribution and a heavy-tailed one, with a dynamic weighting function designed to ensure that the former controls the bulk while the latter controls upper-tail extremes. 
Although conceptually appealing, such approaches are difficult to implement in practice, as the weight function parameters are difficult to estimate. 
Alternatively, \citet{Carreau2008,Carreau2009} proposed to stitch a Pareto distribution with a Gaussian one in order to obtain, within a neural network based approach, a flexible building block that handles both heavy tail behavior and the bulk, but the lower tail behavior was ignored. 
To comply with EVT on both sides of the distribution, \cite{naveau2016modeling} proposed a general class of models that includes simple parametric forms \citep[see, e.g.,][]{Papastathopoulos:Tawn:2013} and more complex semi-parametric models \citep[see, e.g.][]{tencaliec2020}. 
Such models have been used to model precipitation data \citep[see, e.g.,][]{evin2018,Gamet22,haruna2023modeling}, 
 % \cite[see, e.g.][]{taillardat2019,rivoire21,Rivoire22,Legall2022}.
and similar constructions were proposed by \cite{stein2021a,stein2021b}. In this work, we propose and study an extension of the so-called extended generalized Pareto distribution of \cite{naveau2016modeling} to the multivariate case. 
 The main question is thus how to integrate multivariate EVT concepts into this new modeling framework, while keeping a flexible bulk, bypassing threshold selections, and enabling fast inference with our new model. 
%To reach this goal, we need to recall a few elements of this theory. 

Most of the mathematical aspects of multivariate EVT are well established and  can be found in the books of   \cite{beirlant:goegebeur:teugels:segers:2004}
%\cite{kulik_heavy-tailed_2020} 
and \cite{dehaan:ferreira:2006}. 
Various applied treatments are covered by the recent handbook of \cite{Carvalho25}.  
In particular, multivariate threshold exceedances 
are classically defined in terms of a norm and modeled with a multivariate generalized Pareto distribution \citep[see, e.g.,][]{rootzen2018}. 
In this multivariate context, the choice of thresholds to define joint extremes not only remains a hurdle for practitioners, but it is even more complex than the univariate case. 
Furthermore, the computational burden when fitting multivariate generalized Pareto distributions using likelihood-based techniques increases rapidly with the dimension as classical censoring of marginally non-extreme observations make it very challenging. 
These two limitations have led to so-called subasymptotic models and to different inference strategies \cite[see, e.g.,][]{morris2017space,wadsworth2017modelling,Krupskii.Huser:2021,shi2024spatial}. 
 These subasymptotic models seek to capture dependence not only at the highest possible quantiles, but also at lower intermediate levels, with flexible forms of tail decay \cite[see, e.g.,][]{zhang2025}.
 %, ideally, subasymptotic models should contain commonly used asymptotic models as a boundary or limiting case. 
 A recurrent modeling strategy  in many studies  is to decompose the dataset at hand into radial and angular components. 
 The radial part corresponds to a univariate projection, e.g., the $L^1$-norm of all components, and a multivariate extreme event is defined when this radial vector is large. 
 The angular part simply rescales the original dataset by the radial component. This representation leads to many different types of ``scale-mixture'' models.  A key hypothesis  in multivariate EVT is to assume that the angle becomes independent of the radius as the latter becomes large. In such a framework, the interpretation is simple: the radius solely drives the upper tail behavior, while the angle entirely controls the dependence structure. The choice of the radial component (or more general ``risk functionals''; see \citealp{deFondevilleDavison18}) %\cite[and its extension to risk functionals; see][]{deFondeville.Davison:2018}
 %and other aspects like convergence rates \cite[under hidden regular variation; see][]{kulik_heavy-tailed_2020}
 and other considerations (e.g., hidden regular variation; see \citealp{kulik_heavy-tailed_2020}) 
 have led to a plethora of extremal models that can produce different types of extremal dependence structures \cite[see, e.g.,][]{huser2017bridging,engelke2019,wadsworth2022higher,murphy2024inference,Mackay:Jonathan:2024,bacro_multivariate_2024,kakampakou2024spatial,hazra2024efficient}.
In this work, we only focus on the most classical setup, the so-called asymptotic dependence case, which implies that multivariate extremes are strongly concomitant. As we require our multivariate extended generalized Pareto distribution to be compliant with multivariate EVT in both the upper and lower tails, two different angular components are needed. Depending on the value of the radial component, the upper (lower) extremal dependence is captured in our modeling framework by the upper (lower) angular component when the radius becomes large (small).  

Our proposed model differs substantially from alternative approaches for blending the bulk with the tails without any threshold selection. For example, in the spirit of \citet{Frigessi2002}, \citet{VracNaveauBrobinsky07} proposed using a bivariate mixture model with a smooth dynamic weighting function. 
Similarly, \citet{Andr__2024} used a weighted copula mixture model to blend bulk and tails, and \citet{andre2025gaussianmixturecopulasflexible} proposed and studied a different copula model based on a mixture of Gaussian distributions. By contrast, in the present work, we move away from mixture distributions and instead propose a stochastic representation based on a weighted sum that integrates two angular components---one for the upper tail and the one for the lower tail---and where the weight of each term depends on the radial component. In other words, our proposed model is a dynamic mixture performed on the level of the stochastic representation, rather than the distribution or density function.  
Besides facilitating interpretation, another advantage of our construction is that it allows straightforward and fast simulation from the model. This opens the door to simulation-based-inference techniques that avoid costly likelihood-based computations. Precisely, we leverage recent amortized neural inference approaches \citep{Zammit-Mangion.etal:2025}, adapting both neural point estimators and full posterior estimators to our context. More specifically, we deploy neural Bayes estimators \citep{Sainsbury-Dale.etal:2024a} to quickly get point estimates of model parameters, as well as normalizing flows to efficiently approximate the entire posterior distribution while bypassing costly Markov chain Monte Carlo sampling \citep{radev2020bayesflow}. 
 
The paper is organized as follows. After recalling the basics of the univariate extended generalized Pareto distribution, Section \ref{sec: meGPD} details our proposed multivariate  model construction and some of its key properties. Section~\ref{sec:inference} details our neural inference approach, while Section~\ref{sec:simulation} describes a simulation study conducted to compare its performance with more classical estimators. 
In Section~\ref{sec:application}, we illustrate our methodology (both the new model and the neural inference approach) with an application to daily rainfall times series data from the Netherlands over the years 1999--2024. Section~\ref{sec:conclusions} concludes with some discussion.

% 5. \\
% - strategies for inference: hybrid likelihood-moment-based estimators that exploit the model structure, or modern amortized neural inference approaches that rely on simulations from the model\\
% - we propose using neural Bayes estimators and neural posterior approximators, and compare them with a hybrid likelihood-moment-based approach\\
% - we observe gains in efficiency, especially for parameters that are tricky to estimate\\
% 6. \\
% - data application to precipitation or river flow data?\\
% - paper outline\\

%%%%%%%%%%%%%%%%%%%%%%%%%%%%%%%%%
%%%%%%%%%%%%%%%%%%%%%%%%%%%%%%%%%
\section{A multivariate extended generalized Pareto model}
\label{sec: meGPD}

Section~\ref{sec:univariate.eGPD} recalls basics of the univariate extended generalized Pareto distribution; Section~\ref{sec:multiv.eGPD} extends it to the multivariate case, detailing the general construction of our novel multivariate model, and Section~\ref{sec:Properties} studies its marginal and dependence properties in detail.

\subsection{Univariate extended generalized Pareto distribution}\label{sec:univariate.eGPD}
In univariate extreme-value theory, the generalized Pareto distribution (GPD) is commonly used to model high threshold exceedances because it arises as the only possible nondegenerate limiting form, as the threshold tends to the upper endpoint of the distribution. More precisely, let $Y\sim F_Y$ be a random variable with upper endpoint $y_U^\star$. If there exists a rescaling sequence $a_U(u)>0$ such that 
\begin{equation}\label{eq:thr.exc.upper}
\mathbb{P}\!\left(\frac{Y-u}{a_U(u)} \leq y \,\middle|\, Y>u\right)
   \;\to\; H_\xi(y/\sigma), \qquad u \to y_U^\star,
\end{equation}
for some nondegenerate distribution $H_\xi(y/\sigma)$ and for some  positive parameter $\sigma$, then the limit is the GPD with canonical form defined as
\begin{equation}\label{eq:GPD}
H_\xi(y)=\begin{cases}
    1-\left(1+\xi y\right)^{-1/\xi},&\xi\neq0\\
    1-\exp(-y),&\xi=0,
\end{cases}
\end{equation}
defined over the support $\{y>0:1+\xi y>0\}$, where $\xi\in\mathbb{R}$ is a shape parameter, also known as the tail index. This asymptotic result motivates the use of $H_\xi(y/\sigma)$ as a model for unnormalized exceedances $Y-u\mid Y>u$ for high but finite thresholds $u$. Note that the scale parameter $\sigma$ is meant to absorb the sequence $a_U(u)$, and thus may vary with $u$. In particular, if $\xi>0$ and $Y$ has an infinite upper endpoint, i.e., $y_U^\star=\infty$, a simple expansion of \eqref{eq:GPD} yields $\mathbb{P}(Y>y\mid Y>u)\approx (\xi y/\sigma)^{-1/\xi}=K_U y^{-1/\xi}$, as $y\to\infty$, for some constant $K_U>0$. This implies that $Y$ has a heavy, power-law upper-tail behavior.

Although the GPD in \eqref{eq:thr.exc.upper}--\eqref{eq:GPD} characterizes the upper tail of $Y$, it also applies to the lower tail by symmetry. Precisely, if the limit
\begin{equation}\label{eq:thr.exc.lower}
\mathbb{P}\!\left(\frac{v-Y}{a_L(v)} \leq y \,\middle|\, Y<v\right)
   \;\to\; H_{\tilde{\xi}}(y/\tilde{\sigma}), \qquad v \to y_L^\star,
\end{equation}
exists and is nondegenerate, where $a_L(v)$ is another positive rescaling sequence, then $H_{\tilde \xi}(y/\tilde{\sigma})$ is also a GPD of the form \eqref{eq:GPD}, yet with potentially different parameters $\tilde{\sigma}>0,\tilde{\xi}\in\mathbb{R}$. A special case arises when $Y$ is a positive random variable, in which case the lower bound of $Y$ is $y_L^\star=0$. Suppose that for a fixed threshold $v\approx 0$, the distribution of $v-Y\mid Y<v$ is modeled using a GPD with scale $\tilde\sigma>0$ and shape parameter $\tilde\xi<0$. As the upper bound of the support is $-\tilde\sigma/\tilde\xi=v$, this implies that $\tilde\xi=-\tilde\sigma/v$. 
%\pn{No clear to me, how the notation $v$ represents a constant but goes to the lower end point.}
Therefore, in this case, we see that the distribution of low values of $Y$ can be modeled as $Y\mid Y<v\sim (y/v)^{v/\tilde\sigma}\sim K_L y^\kappa$, for some constant $K_L>0$ and lower-tail shape parameter $\kappa=-1/\tilde{\xi}=v/{\tilde\sigma}>0$.

To link both tail behaviors together with a smooth transition in between, one can consider the extended generalized Pareto distribution (eGPD). This name was originally coined by \citet{Papastathopoulos:Tawn:2013} with the objective of  improving upper tail modeling (subasymptotic large extremes). To   integrate the lower tail behavior (subasymptotic low extremes) into the eGPD class,  
\citet{naveau2016modeling} proposed an extension to capture  the full range of nonnegative variables. This updated  eGPD has the general form 
%\begin{equation}\label{eq:eGPD}
$F_Y(y)=B\{H^\kappa_\xi(y/\sigma)\}$, 
%\end{equation}
where $H_\xi(\cdot)$ is the GPD with shape parameter $\xi$ defined in \eqref{eq:GPD}, $\sigma$ is a positive scale parameter, and $B:[0,1]\to[0,1]$ is any continuous cumulative distribution function over the interval $[0,1]$ with positive derivatives at $0$ and $1$, i.e., $B'(0)>0$ and $B'(1)>0$, to make sure that the upper and lower tails are  driven by $\kappa$ and $\xi$, respectively.
The most parsimonious case %\footnote{The function $G(u)=B^\kappa(u)$ was used in \cite{naveau2016modeling}, but this notation was ambiguous, mixing bulk and lower behaviors. The notation with $B(u)$ provides clarity as $\kappa$ is independent of $B(\cdot)$. See \cite{Naveau25} for a recent review on this topic.} 
is to set $B(u)=u$, in which case the eGPD can be expressed as
\begin{equation}\label{eq:eGPD2}
F_Y(y)=\begin{cases}
    \{1-\left(1+\xi y/\sigma\right)^{-1/\xi}\}^\kappa,&\xi\neq0\\
    \{1-\exp(-y/\sigma)\}^\kappa,&\xi=0.
\end{cases}
\end{equation}
When $\xi>0$, it is easy to verify that $1-F_Y(y)\sim\kappa(\xi y/\sigma)^{-1/\xi}$, as $y\to\infty$, and that  $F_Y(y)\sim (y/\sigma)^\kappa$, as $y\to0$. Therefore, this three-parameter eGPD complies with EVT in both tails, with $\kappa$ a shape parameter controlling the lower tail behavior, $\xi$ a shape parameter controlling the upper tail behavior, and $\sigma$ an overall scale parameter affecting the whole distribution from low to high quantiles. While more flexible extensions of this simple eGPD have been proposed by \citet{tencaliec2020} and other types of smooth bulk--tail models have been proposed by 
\citet{stein2021a,stein2021b} and \citet{yadav2021spatial,yadav2022flexible} among others, the simple form \eqref{eq:eGPD2} has found interest in a wide range of environmental applications \citep{Legall2022,yadav2023joint,yadav2025statistics,Haruna24,cisneros2024deep}.

\subsection{General construction of the multivariate eGPD model}\label{sec:multiv.eGPD}
In the sequel, a vector of dimension $d$ is denoted by $\bm x = (x_1,\ldots, x_d)^\top$.
The bold symbol $\bm 0$ refers to a vector all elements of which are equal to 0. Operations between vectors are to
be understood componentwise. For instance, if 
$\bm x = (x_1,\ldots, x_d)^\top$ and $\bm y = (y_1,\ldots, y_d)^\top$,
then $\bm{x} \bm {y}$  a is the vector with components $x_jy_j$, $j=1,\ldots,d$. Similarly, ${\bm x}^{-1}$ is the vector with
components $x_j^{-1}$, $j=1,\ldots,d$.  
The $L^1$ norm of vector $\bm x$  will be denoted by $\|\bm x\|_1=\sum_{j=1}^d |x_j|$.

To extend the eGPD in \eqref{eq:eGPD2} to the multivariate case, we specify its stochastic representation. Having the modeling of precipitation in mind, let $R\sim F_R$ be a positive random variable distributed according to the univariate eGPD \eqref{eq:eGPD2} with $\xi>0$ (i.e., with a heavy upper tail), which will be closely linked to the random vector's marginal behavior. 
To control dependence in the lower and upper tails, respectively, let $\bm L=(L_1,\ldots,L_d)^\top$ and $\bm U=(U_1,\ldots,U_d)^\top$ be (nondegenerate) positive random vectors lying on the $L^1$-sphere (i.e., the ``simplex''), i.e., one has $L_j,U_j\in[0,1]$ such that $\mathbb{P}(L_j>0)=\mathbb{P}(U_j>0)=1$ for all $j=1,\ldots,d$, and $\|\bm L\|_1=\|\bm U\|_1=1$.
%To control dependence in the upper tail, let $\bm U=(U_1,\ldots,U_d)^\top$ be a (nondegenerate) nonnegative random vector lying on the $L^1$-sphere (the ``simplex''), i.e., one has $U_j\geq0$ such that $\mathbb{P}(U_j>0)>0$ for all $j=1,\ldots,d$, and $\|\bm U\|_1=\sum_{j=1}^d U_j=1$. 
%Finally, to control dependence in the lower tail, let $\bm L=(L_1,\ldots,L_d)^\top$ be a positive random vector lying on the ``sphere'' defined by $r(\bm x)=(\sum_{j=1}^d x_j^{-1})^{-1}=\|\bm x^{-1}\|_1^{-1}$, i.e., one has $L_j>0$, $j=1,\ldots,d$, and $r(\bm L)=1$. Note that this implies $L_j\geq1$, $j=1,\ldots,d$. While $r(\cdot)$ is not a norm, it shares the property of being 1-homogeneous, i.e., $r(s\bm l)=s\,r(\bm l)$ for all $s>0$ and $\bm l$.  
%It is thus a valid ``risk functional'' when used in the context of multivariate threshold exceedances modeled through $r$-Pareto distributions under the assumption of multivariate regular variation (see, e.g., \citealp{Dombry.Ribatet:2015,deFondeville.Davison:2018}).
Assume that $R$, $\bm L$, and $\bm U$ are mutually independent. Our proposed multivariate eGPD is then defined through the construction
\begin{equation}\label{eq:meGPD}
\bm Y= R \, \left([1-\omega\{F_R(R)\}]\bm L + \omega\{F_R(R)\}\bm U\right),
\end{equation}
%\begin{equation}\label{eq:meGPD}
%\bm Y= R \, \left([1-\omega\{F_R(R)\}]\bm L^{-1} + \omega\{F_R(R)\}\bm U\right),
%\end{equation}
where  $\omega:[0,1]\to[0,1]$ is a continuous cumulative distribution function defined over the interval $[0,1]$ (i.e., non-decreasing, $\lim_{u\downarrow0}\omega(u)=0$, and $\lim_{u\uparrow1}\omega(u)=1$) controlling the weight attributed to the vectors $\bm L$ (lower tail) and $\bm U$ (upper tail) in the dynamic weighted sum \eqref{eq:meGPD}. Intuitively, as both $\bm L$ and $\bm U$ are on the sphere and the weight function $\omega(\cdot)$ is bounded in $[0,1]$, marginal properties of the random vector $\bm Y$ will be mostly dictated by $R$. Moreover, since $\omega\{F_R(R)\}\approx1$ for large realizations of $R$, the vector $\bm U$ will essentially control the upper tail dependence structure. Conversely, since $\omega\{F_R(R)\}\approx0$ for realizations of $R$ close to zero, the vector $\bm L$ will control the lower tail dependence structure. This is formalized in the next section, where we explore the model properties in more detail. We also note that while we could choose other norms (or even ``risk functionals''; \citealp{deFondevilleDavison18}) than the $L^1$ norm, the model \eqref{eq:meGPD} already provides a fairly flexible framework. 

\subsection{Properties of the multivariate eGPD model}\label{sec:Properties}
\subsubsection{Moments}
While marginal distributions are generally intractable, the marginal moments of each component $Y_j$ (provided they exist) may be expressed from the model definition in \eqref{eq:meGPD} as
%\begin{align}
%\mathbb{E}(Y_j^q)&=\sum_{l=0}^q{q\choose l}\mathbb{E}\{R^q(1-\omega_R)^l\omega_R^{q-l}\}\mathbb{E}(L_j^{-l})\mathbb{E}(U_j^{q-l}),\quad q=1,2,\ldots,\label{eq:meGPDmoments}
%\end{align}
\begin{align}
\mathbb{E}(Y_j^q)
  &= \sum_{l=0}^q \binom{q}{l}\,
     \mathbb{E}\!\left\{ R^q (1-\omega_R)^l \omega_R^{q-l} \right\}
     \mathbb{E}(L_j^{l}) \mathbb{E}(U_j^{q-l}),
     \quad q=1,2,\ldots, \label{eq:meGPDmoments}
\end{align}
where $\omega_R=\omega\{F_R(R)\}$. As $\|\bm L\|_1=\|\bm U\|_1=1$, one always has $\mathbb{E}(L_j^{l})< 1$ and ${\mathbb{E}(U_j^{q-l})<1}$, and the existence of the $q$-th moment in \eqref{eq:meGPDmoments} will therefore depend on the existence of ${\mathbb{E}\{R^q(1-\omega_R)^l\omega_R^{q-l}\}}$, $l=0,\ldots,q$. Since $\omega(u)\in[0,1]$, a sufficient condition for it is the finiteness of $\mathbb{E}(R^q)$, and as $R$ is an eGPD, this is ensured with $\xi<1/q$. Similarly, when $\xi<1/2$, the covariance structure of $\bm Y$ can also straightforwardly be obtained as
%\begin{align}
%\mathbb{C}{\rm ov}(Y_j,Y_k)&=\mathbb{E}(Y_jY_k)-\mathbb{E}(Y_j)\mathbb{E}(Y_k)\nonumber\\
%&=\mathbb{E}\{R^2(1-\omega_R)^2\}\mathbb{E}(L_j^{-1}L_k^{-1})+\mathbb{E}\{R^2(1-\omega_R)\omega_R\}\mathbb{E}(L_j^{-1})\mathbb{E}(U_k)\nonumber\\
%&+\mathbb{E}\{R^2(1-\omega_R)\omega_R\}\mathbb{E}(L_k^{-1})\mathbb{E}(U_j)+\mathbb{E}(R^2\omega_R^2)\mathbb{E}(U_jU_k)\nonumber\\
%&-\mathbb{E}\{R(1-\omega_R)\}^2\mathbb{E}(L_j^{-1})\mathbb{E}(L_k^{-1})-\mathbb{E}\{R(1-\omega_R)\}\mathbb{E}(\omega_R)\mathbb{E}(L_j^{-1})\mathbb{E}(U_k)\nonumber\\
%&-\mathbb{E}\{R(1-\omega_R)\}\mathbb{E}(\omega_R)\mathbb{E}(L_k^{-1})\mathbb{E}(U_j)-\mathbb{E}(R\omega_R)^2\mathbb{E}(U_j)\mathbb{E}(U_k).\label{eq:meGPDcov}
%\end{align}
\begin{align}
\mathbb{C}{\rm ov}(Y_j,Y_k)%&=\mathbb{E}(Y_jY_k)-\mathbb{E}(Y_j)\mathbb{E}(Y_k)\nonumber\\
&=\mathbb{E}\{R^2(1-\omega_R)^2\}\mathbb{E}(L_jL_k)+\mathbb{E}\{R^2(1-\omega_R)\omega_R\}\mathbb{E}(L_j)\mathbb{E}(U_k)\nonumber\\
&+\mathbb{E}\{R^2(1-\omega_R)\omega_R\}\mathbb{E}(L_k)\mathbb{E}(U_j)+\mathbb{E}(R^2\omega_R^2)\mathbb{E}(U_jU_k)\nonumber\\
&-\mathbb{E}\{R(1-\omega_R)\}^2\mathbb{E}(L_j)\mathbb{E}(L_k)-\mathbb{E}\{R(1-\omega_R)\}\mathbb{E}(\omega_R)\mathbb{E}(L_j)\mathbb{E}(U_k)\nonumber\\
&-\mathbb{E}\{R(1-\omega_R)\}\mathbb{E}(\omega_R)\mathbb{E}(L_k)\mathbb{E}(U_j)-\mathbb{E}(R\omega_R)^2\mathbb{E}(U_j)\mathbb{E}(U_k),\;\; j,k=1,\ldots,d.\label{eq:meGPDcov}
\end{align}
As with the marginal moments, the covariance cannot be expressed in closed form in full generality, but it can be easily approximated by numerical simulations. Equations~\eqref{eq:meGPDmoments} and \eqref{eq:meGPDcov} also reveal that all model components, i.e., $R$, $\bm L$ and $\bm U$, jointly affect the marginal moments and covariance behavior of $\bm Y$ in a non-trivial way. 

\subsubsection{Upper and lower tail behaviors}
To understand the (upper or lower) tail structure of $\bm Y$, first note that
%\begin{equation}\label{eq:L1norm}
%\|\bm Y\|_1 = R\left([1-\omega\{F_R(R)\}]\|\bm L^{-1}\|_1 + \omega\{F_R(R)\}\|\bm U\|_1\right)=R,
%\end{equation}
\begin{equation}\label{eq:L1norm}
\|\bm Y\|_1 = R\left([1-\omega\{F_R(R)\}]\|\bm L\|_1 + \omega\{F_R(R)\}\|\bm U\|_1\right)=R,
\end{equation}
given that $\|\bm L\|_1=\|\bm U\|_1=1$ by definition. Therefore, the $L^1$-norm of $\bm Y$ is exactly determined by the random variable $R$. %This can be helpful in some applications, e.g., when interest lies in sum-aggregates of $\bm Y$ rather than individual components. 

To get a first glimpse into the upper and lower tail dependence structure of $\bm Y$, it is natural to consider a ``radial-angular'' decomposition. Treating $\|\bm Y\|_1=R$ as the ``radius'', the ``angle'' is obtained as $\bm Y/\|\bm Y\|_1=[1-\omega\{F_R(R)\}]\bm L + \omega\{F_R(R)\}\bm U$. Hence, since $\omega(u)\downarrow0$ as $u\downarrow0$ and $\omega(u)\uparrow1$ as $u\uparrow1$, and with $R$, $\bm L$ and $\bm U$ being independent, it follows that 
\begin{align*}
\mathbb{P}(\bm Y/\|\bm Y\|_1\leq \bm u\mid \|\bm Y\|_1>r)&\to \mathbb{P}(\bm U\leq \bm u),\qquad r\to\infty;\\
\mathbb{P}(\bm Y/\|\bm Y\|_1\leq \bm l\mid \|\bm Y\|_1<r)&\to \mathbb{P}(\bm L\leq \bm l),\qquad r\to0.
\end{align*}
In other words, the (angular) dependence structure of ``high extremes'' and ``low extremes'' of $\bm Y$, when defined in terms of the $L^1$-norm (being large or small), are completely determined by the vectors $\bm U$ and $\bm L$, respectively.

More detailed theoretical insights can be achieved through the lens of regular variation and related tools, such as Breiman's Lemma \citep{Breiman:1965} and extensions thereof \citep[see, e.g.,][]{Fougeres:Mercadier:2012}. In the univariate case, a random variable $X$ is said to be regularly varying with index $\alpha>0$, scaling sequence $a(t)>0$ and limit measure $\nu$ if 
$t\mathbb{P}(X/a(t)\in \cdot)\stackrel{v}{\to}\nu(\cdot)$, as $t\to\infty$, 
where $\stackrel{v}{\to}$ denotes the vague convergence of measures (i.e., convergence holds for all compact Borel sets with boundary of measure zero). The scaling sequence can always be taken as $a(t)=t^{1/\alpha}$, and because the limit measure $\nu$ is then necessarily homogeneous of order $-\alpha$, i.e., $\nu(t A)=t^{-\alpha}\nu(A)$ for all $t>0$ and Borel sets $A$, the upper tail of $X$ can therefore be characterized as $\mathbb{P}(X>x)\sim\nu((1,\infty))x^{-\alpha}$ as $x\to\infty$. This implies that the random variable $X$ is heavy-tailed with tail index $\xi=1/\alpha$. 

Similarly, in the multivariate case, a random vector $\bm X$ is said to be (multivariate) regularly varying with index $\alpha>0$, scaling sequence $a(t)>0$ and limit measure $\nu$ if 
$t\mathbb{P}(\bm X/a(t)\in \cdot)\stackrel{v}{\to}\nu(\cdot)$, as $t\to\infty$, 
where similar properties hold for $a(t)$ and $\nu$. Intuitively, multivariate regular variation characterizes random vectors that are (both marginally and jointly) heavy-tailed and asymptotically dependent. Furthermore, if $\bm X$ is multivariate regularly varying, then it is in the max-domain of attraction of a max-stable random vector with (heavy-tailed) Fr\'echet marginals \citep[see, e.g.,][for further details on regular variation and extreme-value theory]{Resnick1987}.

\paragraph{Upper tail}
To study the upper tail of $\bm Y$ in our setting, we can equivalently study the upper tail of $R\bm U$ (since $\omega(u)\uparrow1$ as $u\uparrow1$). Note that $R$ follows a heavy-tailed eGPD with $\xi>0$, so it is regularly varying with index $\alpha_R=1/\xi>0$, scaling sequence $a_R(t)=t^\xi$ and limit measure $\nu_R((r,\infty))=\kappa(\sigma/\xi)^{1/\xi}r^{-1/\xi}$ for all $r>0$. Further, given $\|\bm U\|_1=1$, each component $U_j$ automatically meets the condition that there exists $\varepsilon>0$ such that ${0<\mathbb{E}(U_j^{1/\xi+\varepsilon})<\infty}$; in fact, this holds for all $\varepsilon>0$ as $|U_j|\leq 1$. According to Breiman's Lemma \citep[see][]{Breiman:1965}, $RU_j$ is also regularly varying with the same index $\alpha_{RU_j}=\alpha_{R}=1/\xi$ and scaling sequence $a_{RU_j}(t)=a_{R}(t)=t^\xi$, but with a modified limiting measure: $\nu_{RU_j}((r,\infty))=\mathbb{E}(U_j^{1/\xi})\nu_R((r,\infty))$. Thus, the marginal upper tail of $Y_j$ behaves as 
$$\mathbb{P}(Y_j>y)\sim\mathbb{P}(RU_j>y)\sim \mathbb{E}(U_j^{1/\xi})\kappa\left({\sigma/\xi}\right)^{1/\xi}y^{-1/\xi},\qquad y\to\infty.$$
Hence, $Y_j$ is heavy-tailed in the upper tail with the same tail index as the random variable $R$, but its scale is modified through the random variable $U_j$. 

The joint upper tail of the vector $\bm Y$ can similarly be studied through a multivariate extension of Breiman's Lemma. Note that one can write $R\bm U=\bm R\bm U$, where $\bm R=(R,\ldots,R)^\top$ is a fully dependent $d$-dimensional random vector that is multivariate regularly-varying with index $\alpha_{\bm R}=1/\xi$, scaling sequence $a_{\bm R}(t)=t^\xi$ and limit measure $\nu_{\bm R}([\bm 0,\bm r]^C)=\max(r_1^{-1/\xi},\ldots,r_d^{-1/\xi})\kappa (\sigma/\xi)^{1/\xi}$, $\bm r=(r_1,\ldots,r_d)^\top$, with $A^C$ the complement of the set $A$. Thus, thanks to Theorem~3 of \citet{Fougeres:Mercadier:2012}, $\bm R\bm U$ is also multivariate regularly-varying with the same index and scaling sequence, but with limiting measure 
\begin{align*}
\nu_{\bm R\bm U}([\bm 0,\bm r]^C)&=\mathbb{E}\{\nu_{\bm R}(\bm U^{-1}[\bm 0,\bm r]^C)\}=\mathbb{E}\{\nu_{\bm R}([\bm 0,\bm r\bm U^{-1}]^C)\}\\
&=\mathbb{E}\left[\max\left\{\left({r_1/U_1}\right)^{-1/\xi},\ldots,\left({r_d/U_d}\right)^{-1/\xi}\right\}\right]\kappa (\sigma/\xi)^{1/\xi}.
\end{align*}
This implies that the joint upper tail of $\bm Y$ can be characterized by
\begin{align*}
\mathbb{P}(\bm Y\in t[\bm 0,\bm y]^C)&\sim \mathbb{P}(\bm R\bm U\in t[\bm 0,\bm y]^C)\sim \nu_{\bm R\bm U}([\bm 0,t\bm y]^C)\\
&=\mathbb{E}\left[\max\left\{\left({y_1/U_1}\right)^{-1/\xi},\ldots,\left({y_d/U_d}\right)^{-1/\xi}\right\}\right]\kappa (\sigma/\xi)^{1/\xi}t^{-1/\xi},\qquad t\to\infty.
\end{align*}
From this expression, we see how $\bm U$ modulates the joint upper tail structure of $\bm Y$. 

%Another way to see this is to consider the extremal behavior of $\bm Y$ when normalized to the $L^1$-sphere. In other words, one can study the behavior of the ``angular component'' $\bm Y/\|\bm Y\|_1$ as the ``radius'' $\|\bm Y\|_1$ gets large. Since, by construction, we have $\bm Y/\|\bm Y\|_1=[1-\omega\{F_R(R)\}]\bm L + \omega\{F_R(R)\}\bm U$, with $\omega(u)\uparrow1$ as $u\uparrow1$ and $\bm U$ and $R$ being independent, it is easy to see that for all $\bm u$,
%$$\mathbb{P}(\bm Y/\|\bm Y\|_1\leq \bm u\mid \|\bm Y\|_1>r)\to \mathbb{P}(\bm U\leq \bm u),\qquad r\to\infty.$$
%This confirms once more that the joint upper tail behavior is determined solely by the random vector $\bm U$, which controls the (asymptotic) distribution of ``angles'' as the ``radius'' gets large. Furthermore, the angle and the radius are asymptotically independent (as $\|\bm Y\|_1\to\infty$), as required by multivariate regular variation.

\paragraph{Lower tail}
To study the lower tail of $\bm Y$ in our setting, we can similarly study the lower tail of $R\bm L$ (since $\omega(u)\downarrow0$ as $u\downarrow0$) or, equivalently, characterize the upper tail behavior of $R^{-1}\bm L^{-1}$. Note that $R^{-1}$ is regularly-varying with index $\alpha_{R^{-1}}=\kappa$, scaling sequence $a_{R^{-1}}(t)=t^{1/\kappa}$, and limit measure $\nu_{R^{-1}}((r,\infty))=\sigma^{-\kappa}r^{-\kappa}$. Two main cases can arise given that $\bm L^{-1}$ is not necessarily upper-bounded nor lighter tailed than $R^{-1}$: (i) the first case arises when each component $L_j$ meets the condition that there exists $\varepsilon>0$ such that ${\mathbb{E}(L_j^{-(\kappa+\epsilon)})<\infty}$, $j=1,\ldots,d$. This is trivially true when $L_j$ is lower bounded (i.e., ${L_j>\delta>0}$ for some $\delta>0$, almost surely). Then, Breiman's Lemma and its multivariate extension imply that the marginal and joint lower tail of $\bm Y$ behave, respectively, as
\begin{align}
\mathbb{P}(Y_j\leq y)&\sim \mathbb{P}(R^{-1}L_j^{-1}> y^{-1})\sim \mathbb{E}(L_j^{-\kappa})\sigma^{-\kappa}y^\kappa,\quad y\to0;\label{eq:lowertail.mar1}\\
\mathbb{P}(\bm Y \in t[\bm y,\bm\infty)^C)&\sim \mathbb{P}(\bm R^{-1} \bm L^{-1} \in t^{-1}[\bm 0,\bm y^{-1}]^C)\nonumber\\
&\sim \mathbb{E}\left[\max\left\{\left({y_1/L_1}\right)^{\kappa},\ldots,\left({y_d/L_d}\right)^{\kappa}\right\}\right]\sigma^{-\kappa}t^\kappa,\quad t\to0.\label{eq:lowertail.dep1}
\end{align}
Thus, in this case, $Y_j$ has the same lower tail index as $R$ itself, but its scale is modified through the random variable $L_j$. Furthermore, the random vector $\bm L$ modulates the joint lower tail dependence structure of $\bm Y$. (ii) The second case arises when $\bm L^{-1}$ is multivariate regularly varying with index $\alpha_{\bm L^{-1}}>0$ and heavier-tailed than $R^{-1}$, i.e., $\alpha_{\bm L^{-1}}<\kappa$ such that $\mathbb{E}(R^{-\alpha_{\bm L^{-1}}+\varepsilon})<\infty$ for some $\varepsilon>0$. Then, using Breiman's Lemma and its multivariate extension, the marginal and joint lower tail structure of $\bm Y$ can be written, respectively, as 
\begin{align}
\mathbb{P}(Y_j\leq y)&\sim \mathbb{P}(R^{-1}L_j^{-1}> y^{-1})\sim \mathbb{E}(R^{-\alpha_{\bm L^{-1}}})\nu_{L_j^{-1}}((1,\infty))y^{\alpha_{\bm L^{-1}}},\;\; y\to0;\label{eq:lowertail.mar2}\\
\mathbb{P}(\bm Y \in t[\bm y,\bm\infty)^C)&\sim \mathbb{P}(\bm R^{-1} \bm L^{-1} \in t^{-1}[\bm 0,\bm y^{-1}]^C)\sim \mathbb{E}\left\{\nu_{\bm L^{-1}}([\bm 0,R\bm y^{-1}]^C)\right\}t^{\alpha_{\bm L^{-1}}},\;\; t\to0.\label{eq:lowertail.dep2}
\end{align}
where $\nu_{L_j^{-1}}$ and $\nu_{\bm L^{-1}}$ denote the limit measures of $L_j^{-1}$ and $\bm L^{-1}$, respectively. Therefore, in this case, the marginal lower tail of $\bm Y$ is dictated by the tail index of $\bm L^{-1}$, which also controls the lower tail dependence structure through its limit measure.

%To get further insight into the lower joint tail behavior, we can also consider a ``radial-angular'' decomposition of $\bm Y$. Given that, by definition, $\|\bm Y\|_1=R$ and that $\bm Y/\|\bm Y\|_1=[1-\omega\{F_R(R)\}]\bm L^{-1} + \omega\{F_R(R)\}\bm U$, with $\omega(u)\downarrow0$ as $u\downarrow0$ and $\bm L$ and $R$ being independent, it is easy to see that for all $\bm l$,
%\begin{align*}
%\mathbb{P}(\bm Y/\|\bm Y\|_1\leq \bm l\mid \|\bm Y\|_1<r)\to \mathbb{P}(\bm L\leq \bm l),\qquad r\to0.
%\end{align*}
%This implies that the random vector $\bm L$ dictates the (asymptotic) distribution of the ``angles'', $\bm Y/\|\bm Y\|_1$ as the ``radius'' $\|\bm Y\|_1$ gets small. Thus, $\bm L$ solely controls the lower tail dependence structure. As for the upper tail, the angle and the radius are also asymptotically independent (as $\|\bm Y\|_1\to0 $).

% \clearpage
%%%%%%%%%%%%%%%%%%%%%%%%%%%%%%%%%
%%%%%%%%%%%%%%%%%%%%%%%%%%%%%%%%%
\section{Inference}\label{sec:inference}

We now outline several parameter inference methods used with the proposed model. In Section~\ref{sec:likelihoodMOM}, we outline a classical likelihood--method-of-moments approach. In Section~\ref{sec:neural}, we describe two likelihood-free, neural-network-based approaches: neural Bayes estimation \citep[e.g.,][]{Sainsbury-Dale.etal:2024a} and neural posterior estimation \citep[e.g.,][]{radev2020bayesflow}. %In Section~\ref{sec:simulation}, we perform a simulation experiment to assess these inferential approaches. 
Throughout, we assume that each model component has its own set of parameters, denoted by $\bm\theta_R$ for $R$, $\bm\theta_{\bm L}$ for $\bm L$, and $\bm\theta_{\bm U}$ for $\bm U$. The weight function $\omega(\cdot)$ also has its own set of parameters, $\bm\theta_\omega$, and we write $\bm\theta=(\bm\theta_R^\top,\bm\theta_{\bm L}^\top,\bm\theta_{\bm U}^\top,\bm\theta_\omega^\top)^\top$ to denote all model parameters. Let $\bm y_1,\ldots,\bm y_n$ be independent and identically distributed realizations of a random vector $\bm Y$ drawn from the model \eqref{eq:meGPD}. Let $F_{R}$, $F_{\bm L}$, $F_{\bm U}$, and $F_{\bm Y}$ be the cumulative distribution functions corresponding to $R$, $\bm L$, $\bm U$, and $\bm Y$, respectively. Furthermore, assume that all model components admit densities, and denote them by $f_{R}$, $f_{\bm L}$, $f_{\bm U}$, and $f_{\bm Y}$, respectively.

\subsection{Hybrid likelihood--moment estimators}\label{sec:likelihoodMOM}

%Although the full likelihood function involves the intractable data density $f_{\bm Y}$, the latent components $R$, $\bm L$ and $\bm U$ can be indirectly ``observed'' and exploited for inference. 

As a full likelihood estimation will involve the intractable data density $f_{\bm Y}$, the inference presented in this section  has to be hybrid, in the sense that some parameters will be estimated by likelihood maximization while other parameters will be inferred with a method-of-moment approach.  
Hence, this hybrid approach  does not offer all the nice optimality properties of full likelihood estimators. Still, it can be viewed as a benchmark with respect to the neural inference methods presented in 
Section \ref{sec:neural}.

Here, we exploit that the latent components $R$, $\bm L$ and $\bm U$ can be indirectly ``observed''. As noted in \eqref{eq:L1norm}, $\|\bm Y\|_1=R\sim F_R$. 
% Therefore, to estimate $\bm\theta_R$, one can compute the maximum likelihood estimator (MLE) using the observed structure variables $\|\bm y_1\|_1,\ldots,\|\bm y_n\|_1$, that is, by maximizing the following likelihood function:
% $$\mathcal{L}(\bm\theta_R)=\prod_{i=1}^n f_R(\|\bm y_i\|_1;\bm\theta_R).$$
Therefore, the parameter $\bm\theta_R$ can be estimated via maximum likelihood using the statistics $\|\bm y_i\|_1$, $i=1,\ldots,n$, by maximizing the function:
$$\mathcal{L}(\bm\theta_R)=\prod_{i=1}^n f_R(\|\bm y_i\|_1;\bm\theta_R).$$
This yields the estimate $\hat{\bm\theta}_R^{\rm MLE}$.
Next, since $F_{\bm L}$ (or $F_{\bm U}$) characterizes the asymptotic distribution of the angles $\bm Y/\|\bm Y\|_1$ given that the radius $\|\bm Y\|_1$ converges to zero (or infinity), one can set a low threshold $r_0^L$ (or high threshold $r_0^U$), extract low (or high) extremes from the dataset, and maximize the corresponding likelihood function. Specifically, define the index sets $\mathcal{I}_{L}=\{i=1,\ldots,n:\|\bm y_i\|_1<r_0^L\}$ and $\mathcal{I}_{U}=\{i=1,\ldots,n:\|\bm y_i\|_1>r_0^U\}$. One can estimate $\bm\theta_{\bm L}$ and $\bm\theta_{\bm U}$ by maximizing, respectively,
\begin{equation*}
\mathcal{L}(\bm\theta_{\bm L})=\prod_{i\in\mathcal{I}_L} f_{\bm L}(\bm y_i/\|\bm y_i\|_1;\bm\theta_{\bm L}),\qquad\mbox{and}\qquad \mathcal{L}(\bm\theta_{\bm U})=\prod_{i\in\mathcal{I}_U} f_{\bm U}(\bm y_i/\|\bm y_i\|_1;\bm\theta_{\bm U}).
\end{equation*}
This yields the estimates $\hat{\bm\theta}_{\bm L}^{\rm MLE}$ and $\hat{\bm\theta}_{\bm U}^{\rm MLE}$, respectively. Estimating $\bm\theta_\omega$ is more difficult, as the weight function $\omega(\cdot)$ affects the entire distribution from low to high quantiles and there is no general closed-form likelihood function that can be used to estimate it. However, as $\omega(\cdot)$ mostly affects the bulk of the distribution, a method-of-moments approach might be reasonable. As seen in \eqref{eq:meGPDmoments} and \eqref{eq:meGPDcov}, the weight function affects all moments. Estimators based on moments may thus be obtained by solving a system of equations---whose number should equal the number of parameters in $\bm\theta_\omega$---selected among the following ones: 
\begin{align}
{\rm MOM}_q(\bm\theta_{\omega})
  &= \sum_{j=1}^d \left(\mathbb{E}(Y_j^q) - \overline{y^q_{\cdot j}}\right)^2 = 0,
     \qquad q=1,2,\ldots; \label{eq:MOMq} \\
{\rm MOM}_{\rm cov}(\bm\theta_{\omega})
  &= \sum_{j=1}^d \sum_{k=1}^d
     \left(\mathbb{C}{\rm ov}(Y_j,Y_k)
           - \frac{1}{n-1} \sum_{i=1}^n (y_{ij}-\bar{y}_{\cdot j})(y_{ik}-\bar{y}_{\cdot k})
     \right)^2 = 0, \label{eq:MOM2b}
\end{align}
where $y_{ij}$ is the $j$-th component of realization $\bm y_i$, and $\overline{y^q_{\cdot j}}=n^{-1}\sum_{i=1}^n y_{ij}^q$ is the $q$-th sample moment of the $j$-th component (with $\bar{y}_{\cdot j}\equiv \overline{y^1_{\cdot j}}$ for the sample mean). The theoretical counterparts, denoted by $\mathbb{E}(Y_j^q)$ for the marginal moments and $\mathbb{C}{\rm ov}(Y_j,Y_k)$ for the covariances, are computed from \eqref{eq:meGPDmoments} and \eqref{eq:meGPDcov}, respectively, under parameter $\bm\theta_\omega$ (that we wish to estimate) and the true parameters $\bm\theta_R,\bm\theta_{\bm L},\bm\theta_{\bm U}$ (that are fixed when solving \eqref{eq:MOMq} and/or \eqref{eq:MOM2b}). In practice, these true values are unknown and may replaced with their likelihood-based estimates, $\hat{\bm\theta}_R^{\rm MLE},\hat{\bm\theta}_{\bm L}^{\rm MLE},\hat{\bm\theta}_{\bm U}^{\rm MLE}$ obtained in a first step. As the theoretical moments often do not have closed-form expressions, we approximate them by Monte Carlo (i.e., sample moments) using a large simulation sample with parameters fixed to $\hat{\bm\theta}_R^{\rm MLE},\hat{\bm\theta}_{\bm L}^{\rm MLE},\hat{\bm\theta}_{\bm U}^{\rm MLE},\bm\theta_\omega$. This simulated method-of-moments-based approach \citep{McFadden1989} yields the estimate $\hat{\bm\theta}_{\omega}^{\rm MOM}$.

% Note that $\hat{\bm\theta}_R^{\rm MLE}$ only exploits information from a scalar summary (i.e., the sum) of $\bm Y$ and that both $\hat{\bm\theta}_{\bm L}^{\rm MLE}$ and $\hat{\bm\theta}_{\bm U}^{\rm MLE}$ exploit information from one tail only. Note also that $\hat{\bm\theta}_{\bm L}^{\rm MLE}$ and $\hat{\bm\theta}_{\bm U}^{\rm MLE}$ rely on an asymptotic characterization of the joint lower or upper tails, respectively; they will therefore be subject to some bias if the threshold $r_0$ is not set sufficiently low or high, respectively. Furthermore, $\hat{\bm\theta}_{\omega}^{\rm MOM}$ is not based on the likelihood function, but on moments. Therefore, while these are valid ways to estimate model parameters, this hybrid likelihood--method-of-moments approach may waste information, which might result in a loss in efficiency. %This will be further investigated in the simulation study in Section~\ref{sec:simulation}. Nevertheless, these estimators are useful to quickly get a relatively accurate first guess for the parameter values.

While providing valid inference, the estimators described above each rely on partial information: $\hat{\bm\theta}_R^{\rm MLE}$ uses only a scalar summary of $\bm Y$ (i.e., its $L^1$-norm), while $\hat{\bm\theta}_{\bm L}^{\rm MLE}$ and $\hat{\bm\theta}_{\bm U}^{\rm MLE}$ each exploit information from one tail only. 
Further, the latter also rely on an asymptotic characterization of the lower or upper tails, respectively, and they will thus be subject to some bias if the thresholds $r_0^L$ and $r_0^U$ are not set sufficiently low and high, respectively. The estimator $\hat{\bm\theta}_\omega^{\rm MOM}$, meanwhile, is based on moments rather than the likelihood function, and may therefore suffer from reduced statistical efficiency. This motivates the development of alternative inference methods that better utilize the full information contained in the data.

\subsection{Neural estimators}\label{sec:neural}

To improve inference and estimate all parameters simultaneously using the full information available in the data, while also facilitating uncertainty assessment, a modern possible avenue is to use neural simulation-based inference approaches \citep{Zammit-Mangion.etal:2025}. Such approaches are ``likelihood-free'', in the sense that they do not require the computation of any likelihood function. They are also ``amortized'', in the sense that the entire computational cost is encapsulated in the initial training phase: once the neural network is trained using simulated data (usually a few hours), inference with observed data is extremely fast, and the trained neural network can be reused repeatedly with new data from the same model at almost no computational cost (usually a few milliseconds). 

One such approach uses neural networks to approximate Bayes estimators, which are functionals (point summaries) of the posterior distribution. These \textit{neural Bayes estimators} \citep[NBEs; e.g.,][]{Sainsbury-Dale.etal:2024a,Richards.etal:2024,Andre.etal:2025} are presented in Section~\ref{sec:NBE}. Alternatively, one can approximate the entire posterior distribution using \textit{neural posterior estimators} \citep[NPEs; e.g.,][]{radev2020bayesflow}, discussed in Section~\ref{sec:NPE}. 

Before proceeding, we outline some fundamental elements common to both methods. Both approaches involve sampling a large set $\{\bm\theta^{(k)}\}_{k=1}^K$ of parameter vectors from a prior distribution $p(\bm \theta)$, simulating corresponding datasets $\{\bm y_1^{(k)},\ldots,\bm y_n^{(k)}\}_{k=1}^K$ of conditionally independent draws from the model given each $\bm\theta^{(k)}$, and training a neural network using these parameter--data pairs. To enable amortized inference, where the neural network is trained once and then reused with negligible cost for new datasets from the same model, the prior $p(\bm\theta)$ must be vague enough to cover all possible scenarios in which the neural network may be applied. However, as the prior becomes more vague, both the number $K$ of required simulations and the expressiveness needed from the network tend to increase: a balance is therefore necessary between applicability across different datasets and computational efficiency.  

\subsubsection{Neural Bayes estimators}\label{sec:NBE}

An NBE for the parameter $\bm\theta \in \Theta$ of a parametric statistical model (such as our model in \eqref{eq:meGPD}),  
defined on a sample space $\mathcal{S}$, is constructed as follows. Let $\hat{\bm{\theta}}_{\bm\gamma}: \mathcal{S}^n \to \Theta$ denote a neural network parameterized by $\bm\gamma$, that maps $n$ data replicates as input into parameter estimates as output. Note that the neural-network parameters $\bm\gamma$ are often referred to as ``weights'' and ``biases'' in the machine-learning literature. Then, an NBE is obtained by solving %the optimization task 
\begin{equation}
\label{eq:NN}
\hat{\bm\gamma}=\arg\min_{\bm\gamma} \sum_{k=1}^K \ell\{\bm\theta^{(k)},\hat{\bm{\theta}}_{\bm\gamma}(\bm y_1^{(k)},\ldots,\bm y_n^{(k)})\},
\end{equation}
where $\ell(\cdot,\cdot)$ is a user-specified loss function, such as the $L^1$ (absolute) loss  $\ell(\bm\theta,\hat{\bm\theta})=\|\bm\theta-\hat{\bm\theta}\|_1$ or the $L^2$ (quadratic) loss $\ell(\bm\theta,\hat{\bm\theta})=\|\bm\theta-\hat{\bm\theta}\|_2^2$. The process of performing the optimization task \eqref{eq:NN} is referred to as ``training the network'', and this can be done efficiently using back-propagation and stochastic gradient descent. Once trained, the NBE $\hat{\bm{\theta}}_{\hat{\bm\gamma}}(\cdot)$ can be used by ``plugging in'' any realization $\bm y_1,\ldots,\bm y_n$ observed from the same statistical model, yielding point estimates at almost no computational cost.  

Although NBEs leverage deep learning and are sometimes referred to as ``black-box estimators,'' they are firmly grounded in statistical decision theory. % \citep{Sainsbury-Dale.etal:2024a}. 
 Specifically, the objective function in \eqref{eq:NN} is a Monte Carlo approximation of the Bayes risk, 
\begin{equation}
\label{eq:BayesRisk}
r(\hat{\bm\theta}(\cdot)) = \int_\Theta \bigg\{ \int_{\mathcal{S}^n} \ell\big(\bm\theta, \hat{\bm\theta}(\bm y_1, \ldots, \bm y_n)\big) \prod_{i=1}^n f_{\bm Y}(\bm y_i \mid \bm\theta) \, {\rm d}\bm y_1 \cdots {\rm d}\bm y_n \bigg\} p(\bm\theta) \, {\rm d}\bm\theta,
% r(\hat{\bm\theta}(\cdot)) = \int_\Theta \bigg\{ \int_{\mathcal{S}} \cdots \int_{\mathcal{S}} \ell\big(\bm\theta, \hat{\bm\theta}(\bm y_1, \ldots, \bm y_n)\big) \prod_{i=1}^n f_{\bm Y}(\bm y_i \mid \bm\theta) \, {\rm d}\bm y_1 \cdots {\rm d}\bm y_n \bigg\} p(\bm\theta) \, {\rm d}\bm\theta,
\end{equation}
and NBEs therefore approximate Bayes estimators (minimizers of \eqref{eq:BayesRisk}), which lends them a principled interpretation. In particular, Bayes estimators are functionals of the posterior distribution induced by the prior $p(\bm\theta)$, with the specific functional determined by the choice of loss function $\ell(\cdot,\cdot)$. For instance, the $L^2$ and $L^1$ loss functions yield the posterior mean and the vector of marginal posterior medians, respectively. These estimators enjoy appealing asymptotic properties under mild conditions (e.g., consistency, efficiency; see \citealp{Casella_2001}). Moreover, since neural networks are universal function approximators \citep{Hornik_1989_FNN_universal_approximation_theorem}, an NBE based on a sufficiently flexible neural network can approximate the true Bayes estimator arbitrarily well as the number $K$ of simulations in \eqref{eq:NN} increases; see \citet{rodder2025theoretical} for theoretical results of consistency under regularity conditions and the $L^2$ loss. Therefore, it is unsurprising that NBEs have demonstrated strong empirical performance in practice \citep[e.g.,][]{Sainsbury-Dale.etal:2024a,Andre.etal:2025}. 

Choosing an appropriate architecture (i.e., functional form) for the neural network is key and strongly depends on the structure of the data. In our case, we are interested in making inference for the multivariate model \eqref{eq:meGPD} observed with multiple independent replicates. Following \citet{Sainsbury-Dale.etal:2024a}, we use the DeepSets architecture \citep{Zaheer_2017_Deep_Sets}, which appropriately handles exchangeable data by imposing permutation invariance of the replicates. This architecture may be written as
\begin{equation}\label{eqn:DeepSet:NBE}
\hat{\bm{\theta}}_{\bm\gamma}(\bm y_1,\ldots,\bm y_n)
  = \bm\phi\!\left\{\frac{1}{n}\sum_{i=1}^n \bm\psi(\bm y_i), \, \log(n)\right\},
\end{equation}
where $\bm\phi$ and $\bm\psi$ are neural networks themselves (with $\bm\gamma$ combining their parameters). 
 The role of the inner neural network $\bm\psi$ is to (automatically) extract informative summary statistics, which are then optimally mapped, together with the sample size $n$ on log-scale, to parameter estimates via the outer neural network $\bm\phi$. In the multivariate case, since there is a priori no spatial, temporal, or graphical structure in the data that can be exploited, it is natural to choose both $\bm\phi$ and $\bm\psi$ as standard dense multilayer perceptrons (MLPs) as in \citet{Andre.etal:2025}, \citet{coloma2025fastlikelihoodfreeparameterestimation}, and \citet[][Section~S6]{Sainsbury-Dale_2025_incomplete_data}. The DeepSets architecture can accommodate datasets of different sizes, $n$, but the NBE may not perform well with data that have not been ``seen'' during training. Thus, to make the NBE applicable to varying sample sizes, it is important to randomize the sample size $n$ during training alongside the model parameters and the data. This slightly changes the objective functions in \eqref{eq:NN} and \eqref{eq:BayesRisk}, but the principle remains the same.

Uncertainty assessment with NBEs proceeds naturally using bootstrap techniques \citep[e.g.,][]{lenzi2023neural,Richards.etal:2024,Sainsbury-Dale.etal:2024a}, which are particularly attractive when nonparametric bootstrap is possible (e.g., when the data consist of independent replicates), or when simulation from the fitted model is fast, in which case parametric bootstrap is also computationally efficient. Alternatively, one may construct an NBE that approximates a set of marginal posterior quantiles, which can then be used to construct credible intervals for each parameter \citep{Sainsbury-Dale.etal:2024b,Andre.etal:2025}. Inference then remains fully amortized since, once the NBEs are trained, both point estimates and credible intervals can be obtained with virtually zero computational cost. 

\subsubsection{Neural posterior estimators}\label{sec:NPE}

We now describe NPEs, which allow for amortized approximate posterior inference through the minimization of an expected Kullback--Leibler (KL) divergence \citep{Kullback_1951}. Throughout, $q(\bm \theta; \bm \zeta)$ denotes a parametric approximation to the posterior distribution of $\bm \theta$, where the approximate-distribution parameters $\bm \zeta$ belong to a space $\mathcal{Z}$.

%First, consider the case that inference is based on 
If inference is required for a single dataset $\bm y_1, \dots, \bm y_n$ only, 
%. In this non-amortized setting, 
the parameters $\bm \zeta$ can be chosen to minimize the KL divergence between $p(\bm \theta \mid \bm y_1, \dots, \bm y_n)$ and $q(\bm \theta; \bm \zeta)$: 
\begin{align}
  \hat{\bm \zeta} 
  &= \underset{\bm \zeta}{\mathrm{arg\,min\;}} \mathrm{KL}\big\{p(\bm \theta \mid \bm y_1, \dots, \bm y_n) \,\|\, q(\bm \theta ; \bm \zeta)\big\} \notag \\
  &= \underset{\bm \zeta}{\mathrm{arg\,max\;}} \int_{\Theta} \log\big\{ q(\bm \theta ; \bm \zeta) \big\} \, p(\bm \theta \mid \bm y_1, \dots, \bm y_n) \, \mathrm{d}\bm \theta.
  \label{eqn:kappa_non-amortized}
\end{align}
However, solving this optimization problem is often computationally demanding even for a single dataset. To circumvent this challenge and amortize our inference framework, one may construct a mapping $\hat{\bm \zeta} : \mathcal{S} \to \mathcal{Z}$ that transforms data into the optimal parameters, which can be obtained by solving 
\begin{align}
  \hat{\bm \zeta}(\cdot) 
  &= \underset{\bm \zeta(\cdot)}{\mathrm{arg\,min\;}} \, \mathbb{E}_{\bm Y_1, \ldots, \bm Y_n} \mathrm{KL} \left\{ p(\bm \theta \mid \bm Y_1, \ldots, \bm Y_n) \,\|\, q\big(\bm \theta ; \bm \zeta(\bm Y_1, \ldots, \bm Y_n)\big) \right\}  \notag \\
  &= \underset{\bm \zeta(\cdot)}{\mathrm{arg\,max\;}}  \int_{\mathcal{S}^n} \int_{\Theta} \log \big\{ q(\bm \theta ; \bm \zeta(\bm y_1, \ldots, \bm y_n)) p(\bm y_1, \ldots, \bm y_n \mid \bm \theta)  p(\bm \theta)  \mathrm{d}\bm \theta \bigg\}  \mathrm{d}\bm y_1 \cdots \mathrm{d}\bm y_n.
  \label{eqn:kappa_hat}
\end{align}
This mapping can be modeled using a neural network $\hat{\bm \zeta}_{\bm \gamma} : \mathcal{S} \to \mathcal{Z}$ parameterized by $\bm \gamma$, trained by minimizing a Monte Carlo approximation of the expected KL divergence in \eqref{eqn:kappa_hat}: 
\begin{equation}\label{eqn:kappa_hat_gamma}
\hat{\bm \gamma} = \underset{\bm \gamma}{\mathrm{arg\,max}} \sum_{k=1}^K \log q(\bm \theta^{(k)}; \hat{\bm \zeta}_{\bm \gamma}(\bm y_1^{(k)},\ldots,\bm y_n^{(k)})). 
\end{equation}
Once trained, the neural network $\hat{\bm \zeta}_{\hat{\bm \gamma}}(\cdot)$ can be used to estimate \eqref{eqn:kappa_non-amortized} given an observed dataset at almost no computational cost. The neural network $\hat{\bm \zeta}_{\hat{\bm \gamma}}(\cdot)$ and the corresponding approximate distribution $q(\cdot; \bm \zeta)$ are collectively referred to as an NPE. 

There are numerous choices for the approximate distribution $q(\cdot; \bm \zeta)$. A simple option is a multivariate Gaussian distribution \citep[e.g.,][]{Chan_2018}, where $\bm \zeta$ comprises the mean vector and the non-zero entries of the lower Cholesky factor of the covariance matrix. More flexible alternatives include Gaussian mixtures \citep[e.g.,][]{Papamakarios_Murray_2016} and trans-Gaussian distributions \citep[e.g.,][]{Maceda_2024}. However, the most widely used approach is to adopt normalizing flows for their ability to approximate essentially any arbitrary density functions \citep[e.g.,][]{Ardizzone_2018, Kobyzev_2020, radev2020bayesflow, Papamakarios_2021_review}. A normalizing flow defines a flexible distribution by transforming a simple base distribution (e.g., standard Gaussian) via a sequence of invertible and differentiable mappings, chosen so that the resulting density remains tractable via the change-of-variables formula. This enables efficient density evaluation during training (e.g., when optimizing~\eqref{eqn:kappa_hat_gamma}) and efficient sampling at inference time. A particularly popular and expressive class is the so-called affine coupling flow \citep[e.g.,][]{Dinh_2016, Kingma_Dhariwal_2018, Ardizzone_2019_conditional_normalising_flows}, which is a universal density approximator \citep{Teshima_2020} and is the approximate distribution adopted in this work. 

Finally, selecting an appropriate architecture, that is, the functional form of the neural network $\hat{\bm \zeta}_{\bm \gamma}(\cdot)$, is crucial and should reflect the structure of the data. Since the data structure remains unchanged whether constructing an NBE or NPE, the same architecture can be employed in both cases, with the only difference being the output space of the network. In our setting, %where inference is required for the multivariate model \eqref{eq:meGPD} based on multiple independent replicates, 
we adopt the DeepSets architecture:
\begin{equation}\label{eqn:DeepSet:NPE}
\hat{\bm{\zeta}}_{\bm\gamma}(\bm y_1,\ldots,\bm y_n)
  = \bm\phi\!\left\{\frac{1}{n}\sum_{i=1}^n \bm\psi(\bm y_i), \, \log(n)\right\},
\end{equation}
where $\bm\phi$ and $\bm\psi$ are dense MLPs (with $\bm\gamma$ combining their parameters), and the output space of $\bm\phi$ is the space $\mathcal{Z}$ of approximation-distribution parameters.

\section{Simulation study}\label{sec:simulation}

To validate the estimators described in Section~\ref{sec:inference}, we now conduct a simulation study with data generated from our proposed model~\eqref{eq:meGPD}. Section~\ref{sec:simsetting} first details the setting of our simulation study, and Section~\ref{sec:simresults} discusses the results.

\subsection{Setting}\label{sec:simsetting}

\begin{figure}[t!]
\centering
\includegraphics[width=0.92\linewidth]{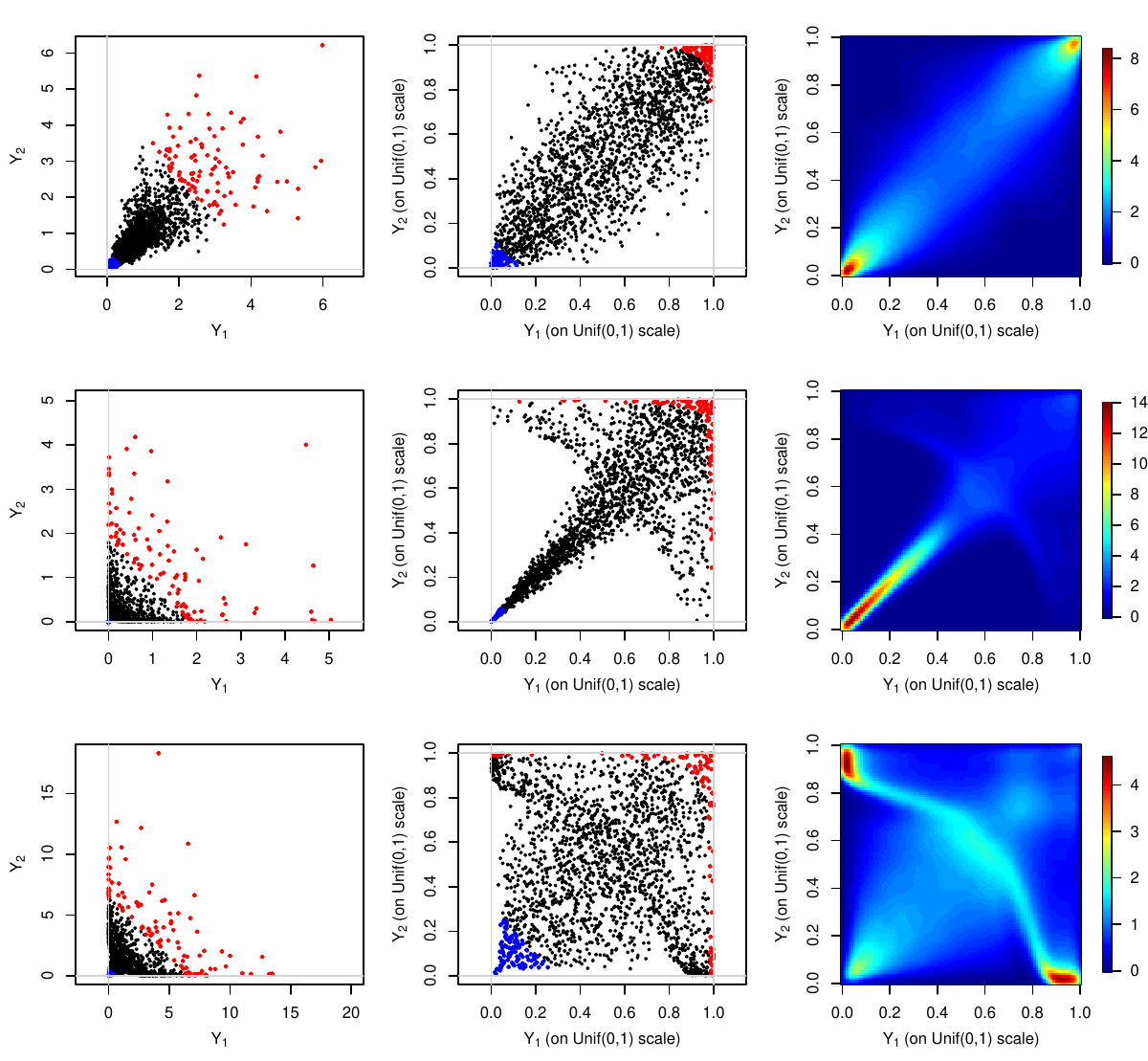}
\caption{Sample of size $n=2000$ from the multivariate eGPD model \eqref{eq:meGPD} with the parametric specification specified in Section~\ref{sec:simsetting}, with $\bm\theta=(\kappa,\sigma,\xi,\theta_{\bm L},\theta_{\bm U},\theta_{\omega})^\top$ set to $(3,1,0.05,10,20,0.25)^\top$ (top), $(0.3,1,0.05,10,0.5,0.25)^\top$ (middle row), and $(3,1,0.2,4,0.5,0.25)^\top$ (bottom), plotted on the original marginal scale (left), or rank-transformed to the ${\rm Unif}(0,1)$ scale (middle column) with a corresponding bivariate kernel density estimate (right). Red and blue points in the first two columns highlight ``high extremes'' (here, the top $5\%$ of $\|\bm Y\|_1$ values) and ``low extremes'' (here, the lowest $5\%$ of $\|\bm Y\|_1$ values).\label{fig:illustration}}
\end{figure}

To assess our methods, we consider the bivariate case with the below model specifications:
\begin{itemize}
\item $R$ follows the univariate eGPD in \eqref{eq:eGPD2} with parameters $\bm\theta_R=(\kappa,\sigma,\xi)^\top\in(0,\infty)^3$;
\item $\bm L=(V_1,1-V_1)^\top$ is defined in terms of the random variable $V_1$ assumed to follow the symmetric Beta$(\theta_{\bm L},\theta_{\bm L})$ distribution, where $\theta_{\bm L}>0$ is a shape parameter controlling the strength of dependence in the lower tail. This ensures that $\|\bm L\|_1=1$ as required. Note that, with this specification, the moment condition $\mathbb{E}(L_j^{-\kappa})<\infty$ is satisfied if and only if $\theta_{\bm L}>\kappa$; in this case, the lower tail structure of $\bm Y$ is determined by \eqref{eq:lowertail.mar1}--\eqref{eq:lowertail.dep1}. When $\theta_{\bm L}<\kappa$, it is instead determined by \eqref{eq:lowertail.mar2}--\eqref{eq:lowertail.dep2}. We can further show that in our setting, $\bm L^{-1}$ is multivariate regularly varying with index $\alpha_{\bm L^{-1}}=\theta_{\bm L}$ and limit measure $\nu_{\bm L^{-1}}([\bm 0,\bm y]^C)
%=\Gamma(2\theta_{\bm L})\{\Gamma(\theta_{\bm L})\}^{-2}\theta_{\bm L}^{-1}
=\nu_{L_1^{-1}}((y_1,\infty))+\nu_{L_2^{-1}}((y_2,\infty))\propto(y_1^{-\theta_{\bm L}}+y_2^{-\theta_{\bm L}})$ being the sum of marginal limit measures, which corresponds to asymptotic independence of the components $L_1^{-1}$ and $L_2^{-1}$. To avoid this case and make sure the moment condition is always satisfied, one could also shift and rescale the Beta distribution of $V_1$ to the support $[\epsilon,1-\epsilon]$ for some small $\epsilon>0$, but we do not investigate this possibility here;
\item $\bm U=(U_1,1-U_1)^\top$ is defined in terms of the random variable $U_1$ assumed to follow the symmetric Beta$(\theta_{\bm U},\theta_{\bm U})$ distribution, where $\theta_{\bm U}>0$ is a shape parameter controlling the strength of dependence in the upper tail. This ensures that $\|\bm U\|_1=1$ as required;
% \item $\omega(u)$, $u\in[0,1]$, is the Beta$(\theta_{\omega},\theta_{\omega})$ cumulative distribution function, where $\theta_{\omega}>0$ is a shape parameter controlling the mixing and speed of transition between the lower and the upper tails.
\item $\omega(u)$, $u \in [0,1]$, is a Beta$(3,3)$ cumulative distribution function, taking as input the argument $(u - \theta_{\omega}) / (1 - 2\theta_{\omega})$, where $\theta_{\omega} \in (0, 0.5)$ is a parameter controlling the transition region between the lower and upper tails.
\end{itemize}
Under this model specification, the parameters of the model are $\bm\theta
% =(\bm\theta_R^\top,\theta_{\bm L},\theta_{\bm U},\theta_{\omega})^\top
=(\kappa,\sigma,\xi,\theta_{\bm L},\theta_{\bm U},\theta_{\omega})^\top$ in the space $\Theta=(0,\infty)^5\times(0, 0.5)$. Figure~\ref{fig:illustration} displays simulations from the model for various choices of parameter values for which the moment condition $\mathbb{E}(L_j^{-\kappa})<\infty$ is satisfied. These plots illustrate the model's flexibility, and its ability to separately control the lower and upper tail structures through the parameters $\kappa,\theta_{\bm L}$ (lower tail) and $\xi,\theta_{\bm U}$ (upper tail).

To infer the model parameters, we use the hybrid likelihood--moment estimators (using the covariances as in \eqref{eq:MOM2b} to estimate $\theta_\omega$), as well as neural estimators. Specifically, we train an ensemble \citep{Sainsbury-Dale_2025_incomplete_data} of five NBEs (see Section~\ref{sec:NBE}) under the $L^1$ loss, \mbox{$\ell(\bm\theta, \hat{\bm\theta}) = \|\bm\theta - \hat{\bm\theta}\|_1$}, along with a single NPE (Section~\ref{sec:NPE}). The neural networks used in both \eqref{eqn:DeepSet:NBE} and \eqref{eqn:DeepSet:NPE} follow the DeepSets architecture, where the functions $\bm\phi$ and $\bm\psi$ are MLPs with three hidden layers of width 128 and ReLU activation functions. Before passing data through the neural networks, we apply a variance-stabilizing transformation, $h(x) = \operatorname{sign}(x)\log(1 + |x|)-1$.  For posterior inference with the NPE, we use a normalizing flow based on affine coupling blocks \citep[e.g.,][]{Dinh_2016, Kingma_Dhariwal_2018, Ardizzone_2019_conditional_normalising_flows}, with default hyperparameters as documented in our chosen software.  We train the neural networks by sampling $K=10^5$ parameter vectors from the following marginal priors, chosen to be relatively vague: \mbox{$\kappa\sim{\rm Unif}(0.1,10)$}, $\sigma\sim{\rm Unif}(0.1,3)$, $\xi\sim{\rm Unif}(0,0.5)$, $\theta_{\bm L}\sim{\rm Unif}(0.1,20)$, $\theta_{\bm U}\sim{\rm Unif}(0.1,20)$, and $\theta_\omega\sim{\rm Unif}(0, 0.5)$. We then generate $K$ sample sizes from the discrete uniform distribution, $n\sim{\rm Unif}\{1000,\ldots,4000\}$, and, conditionally on each parameter-vector and sample-size pair, we simulate a dataset from the model~\eqref{eq:meGPD}. For the validation set used to monitor convergence during training, we generate $K/10$ parameter-data pairs in a similar manner. We cease training when the objective function in \eqref{eq:NN} or \eqref{eqn:kappa_hat_gamma} has not decreased in ten consecutive epochs, where an epoch is defined to be one complete pass through the entire training dataset when doing stochastic gradient descent. %to decrease the objective functions. 

Our experiments are implemented using the package \textbf{NeuralEstimators} \citep{NeuralEstimators}, which is available in \texttt{Julia} and \texttt{R}. We use a workstation with an AMD Ryzen Threadripper Pro 5995WX CPU with 64 cores and 256 GB of RAM, and a NVIDIA RTX A4500 GPU with 21 GB of RAM. All results presented in this section are reproducible via the code available at \url{https://github.com/alotaibin/multivariateEGPD}. 

With these specifications, the total training time for the NBEs and NPE are 17.9 and 11.3~hours, respectively. While this can seem substantial, recall that these neural inference methods are amortized, meaning that the computational cost can be offset by applying the neural network repeatedly at the inference stage, almost instantaneously. 

\subsection{Results}\label{sec:simresults}

\begin{figure}[t!]
\centering
\includegraphics[width=\linewidth]{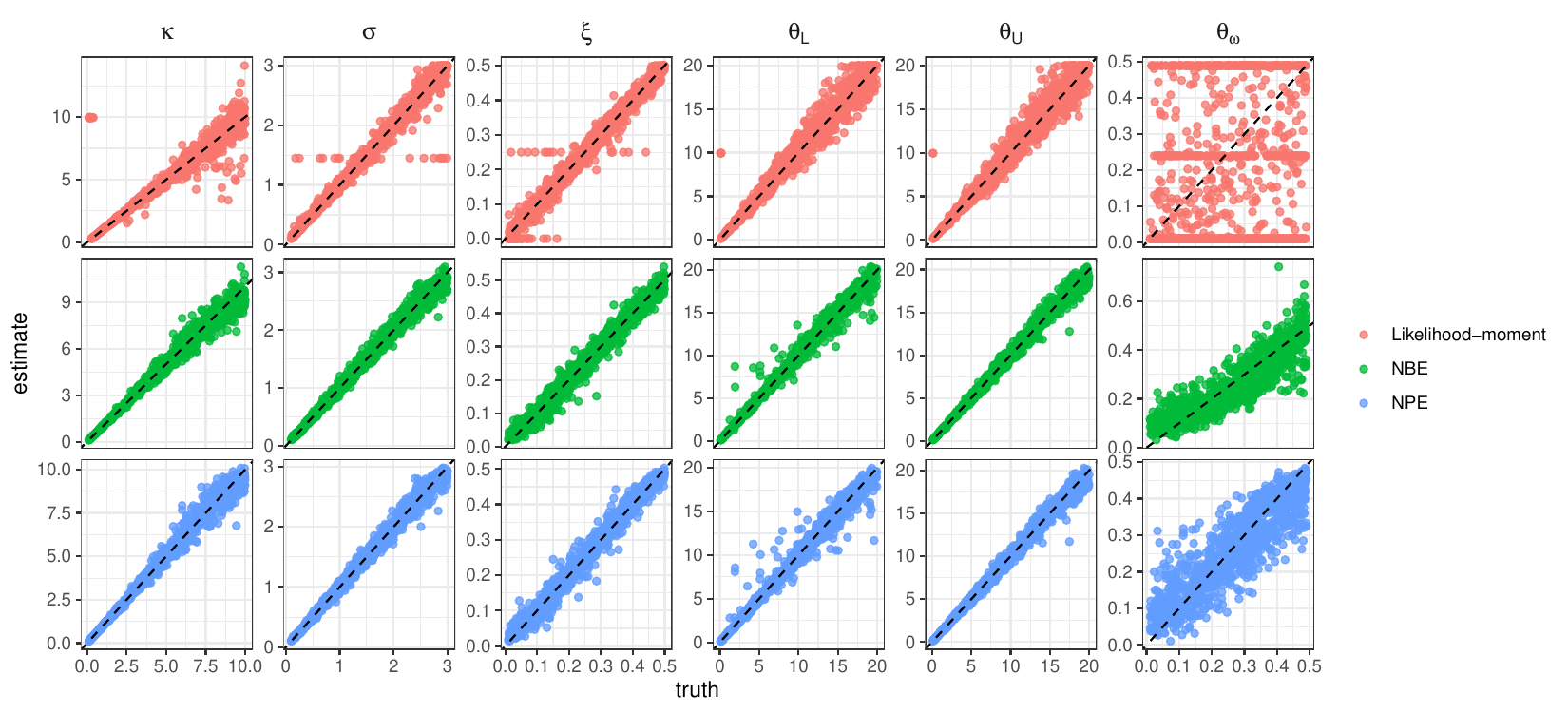}
\caption{Recovery plots for the likelihood-moment-based estimator (top), NBEs (middle), and posterior medians obtained from the NPE (bottom), for the parameters $\kappa$, $\sigma$, $\xi$, $\theta_{\bm L}$, $\theta_{\bm U}$, and $\theta_\omega$ (left to right), based on test data sampled from model~\eqref{eq:meGPD} under the specification described in Section~\ref{sec:simsetting}. Each plot displays parameter estimates against the true values used to generate the test data, with a fixed sample size of $n = 4000$ per dataset.
\label{fig:estimators_assessment}
}
\end{figure}

To assess the estimators, we use a test set of $1000$ parameter-data samples with a fixed sample size of $n=4000$. Figure~\ref{fig:estimators_assessment} presents recovery plots for the likelihood--moment-based estimator (top), the NBEs (middle), and the posterior medians obtained from the NPE (bottom). While all estimators generally yield sensible results, the likelihood--moment-based estimators exhibit higher overall variability compared to the neural approaches. The benefits of neural estimators are particularly evident for the more challenging parameters, namely, the weight function parameter $\theta_\omega$ and the tail parameters $\theta_{\bm L}$ and $\theta_{\bm U}$. Overall, both classes of neural estimators perform well in terms of statistical efficiency. In terms of computational efficiency, the neural methods offer substantial speedups once trained. For a single dataset with $n = 4000$ observations, inference using the NBEs takes 0.004 seconds, the NPE takes 0.007 seconds, whereas the likelihood--moment-based estimator requires 4.2 seconds.

\begin{figure}[t!]
\centering
\includegraphics[width=\linewidth]{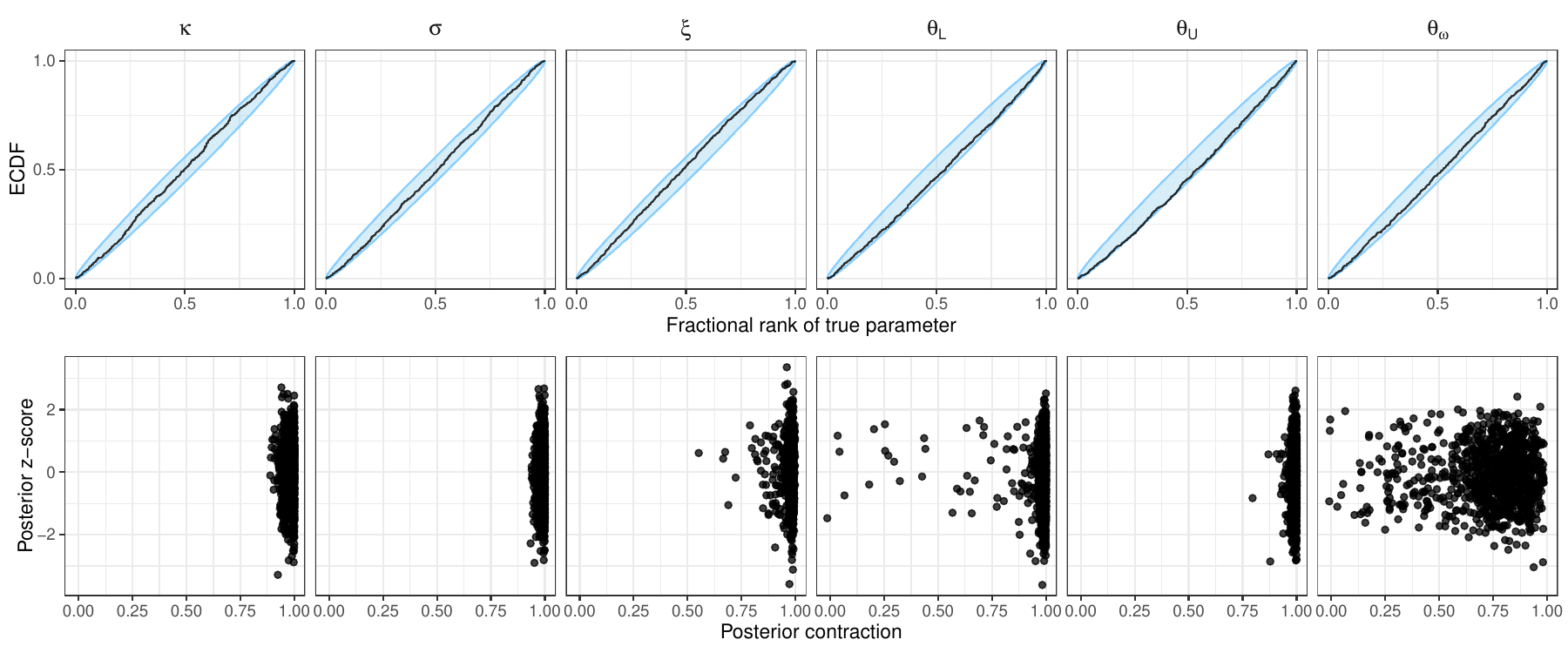}
\includegraphics[width=\linewidth]{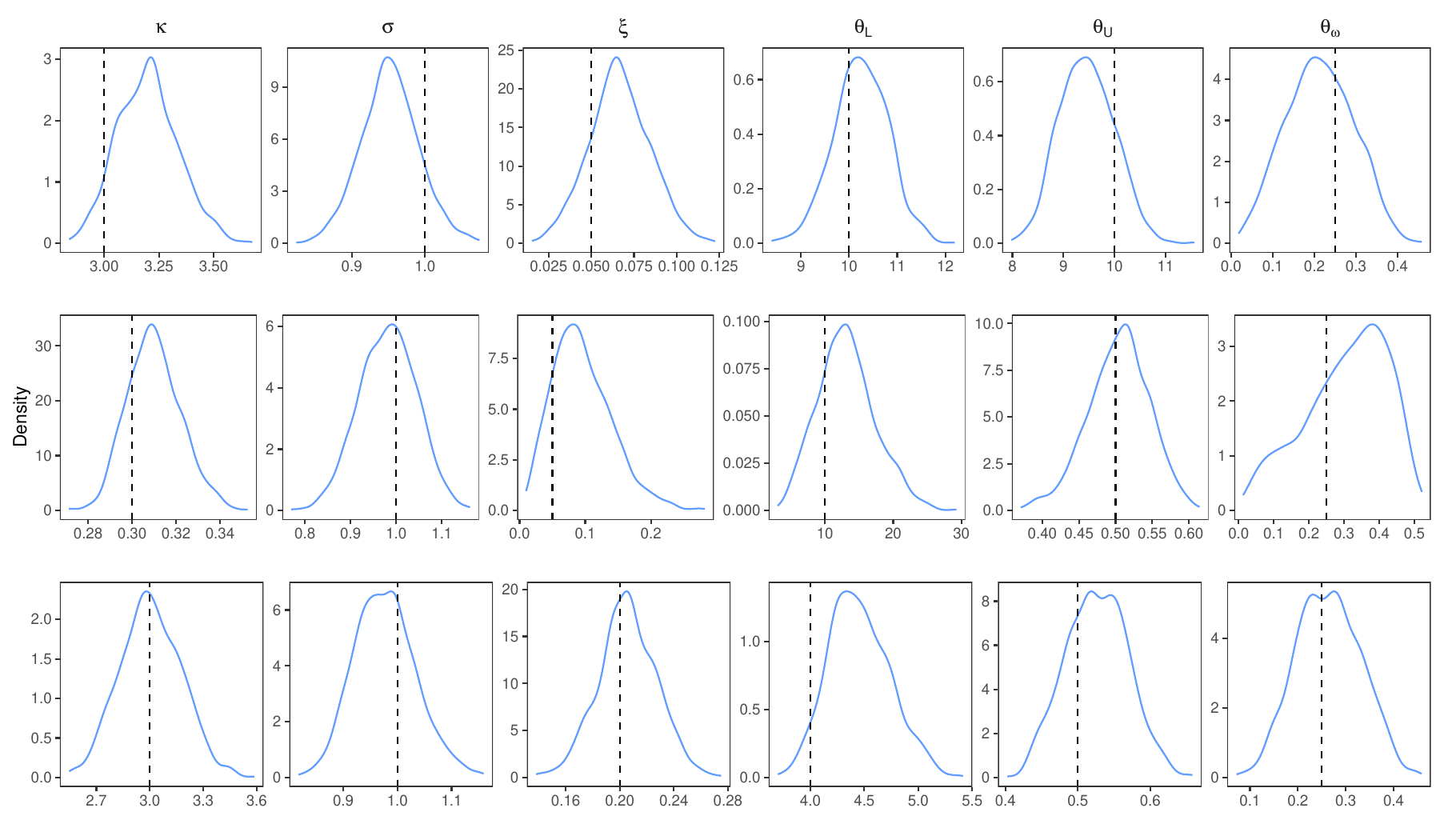}
\caption{
Diagnostic plots assessing the quality of the approximate posterior distributions obtained from the NPE. (Top row) ECDF versions of posterior-prior rank histogram plots, with shaded confidence bands. For well-calibrated posteriors, the curves should lie within the confidence bands. (Second row) Posterior $z$-scores against posterior contractions for each parameter. (Remaining rows) Approximate posterior distributions (blue curves) for each model parameter across three test cases (top to bottom), for three test cases with true parameter values (dashed black vertical lines) corresponding to those displayed in Figure~\ref{fig:illustration}.} \label{fig:estimators_assessment2}
\end{figure}

To better evaluate the accuracy of the approximate posterior distributions obtained from the NPE, we employ several diagnostic tools commonly used in simulation-based inference. Figure~\ref{fig:estimators_assessment2} presents a suite of such diagnostic plots. The top row shows an empirical cumulative distribution function (ECDF) version of posterior-prior rank histograms, also known as simulation-based calibration (SBC) plots \citep{Talts_2018, Sailynoja_2022, Modrak_2025}. If the posterior distributions are calibrated, the ECDF curves should lie within the confidence bands and closely follow the diagonal. This is indeed the case in our results, suggesting that the NPE posterior approximations are well-calibrated across the test set.

The second row of Figure~\ref{fig:estimators_assessment2} plots the posterior $z$-scores against posterior contractions \citep{Schad_2019}. The posterior $z$-score is defined as the difference between the posterior mean and the true parameter value, divided by the posterior standard deviation. The posterior contraction is defined as $1 - \text{var}(\theta \mid \bm Y) / \text{var}(\theta)$ for a given parameter $\theta$, and reflects how much information has been gained from the data. In most cases, a well-calibrated posterior should lead to posterior $z$-scores  that are centered around zero with unit variance, while posterior contraction values close to 1 indicate strong learning from the data. In our results, the posterior $z$-scores show no systematic bias for any parameter, and the posterior contractions are high for most parameters. However, for the more challenging parameter $\theta_\omega$, contractions are occasionally lower, reflecting inherent difficulty in estimating this component reliably. 

The remaining rows of Figure~\ref{fig:estimators_assessment2} show the full approximate posterior distributions for each parameter across three test cases previously shown in Figure~\ref{fig:illustration}. Across all cases, the true parameter values (shown as vertical dashed lines) lie within the support of the approximate posterior distributions. Collectively, these diagnostics provide empirical support that the NPE approximates the true posterior well across the test datasets. 

% Other useful resources:
% https://betanalpha.github.io/assets/case_studies/principled_bayesian_workflow.html#112_Prior_Pushforward_Checks
% https://hyunjimoon.github.io/SBC/articles/rank_visualizations.html

%Estimates obtained from NBEs and the likelihood-moment-based estimators are also displayed for comparison. 
%Overall, full posteriors obtained from the NPEs are aligned with point estimates obtained from the NBEs, though some disagreement can be spotted in a few cases. %As noted above, neural methods based on NPEs and NBEs clearly outperform the likelihood-moment-based estimators, which can be quite unstable in some cases, especially with $\kappa$ and $\sigma$ in the second test case (middle row).

\section{Analysis of three Dutch precipitation time series}\label{sec:application}
%to precipitation data}

%\subsection{Data and background}

\begin{figure}[t!]
  \centering
  \includegraphics[width=0.85\linewidth]{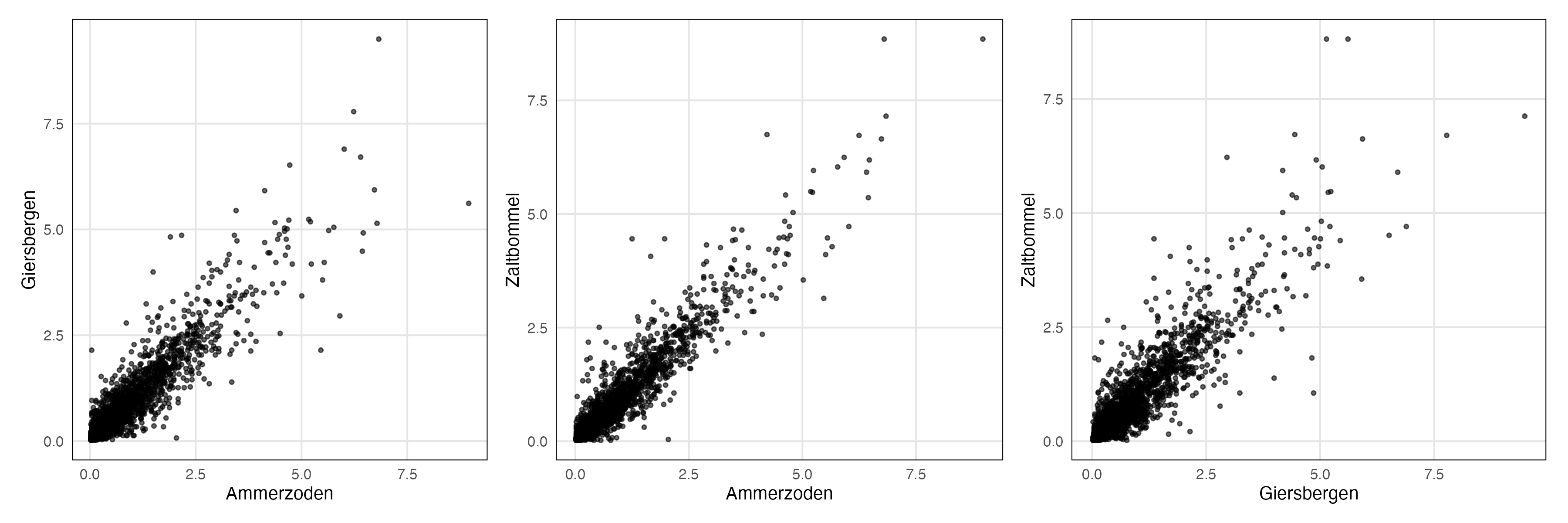}
  \caption{Pairwise scatterplots of fall–winter daily scaled precipitation for the three station pairs: Ammerzoden–Giersbergen (left), Ammerzoden–Zaltbommel (center) and Giersbergen–Zaltbommel (right).}
  \label{fig:scatter_precip}
\end{figure}

Modeling low, moderate and extreme rainfall jointly across multiple locations is useful for regional flood-risk assessment and planning: severe floods can arise from single heavy storms or from accumulations of more moderate rainfall events. To illustrate our multivariate eGPD, we use daily precipitation totals (mm) from three North Brabant meteorological stations in the Netherlands: Ammerzoden, Giersbergen, and Zaltbommel (see Table~\ref{tab:station_info}) obtained from the European Climate Assessment \& Dataset (ECA\&D; \citealp{tank2002publications}). Following \citet{Bortot.Gaetan:2022}, we retain only the fall--winter months (October--February) over 1999--2024 and eliminate all zero-rain days, producing $n=2\,138$, $2\,166$, and $2\,133$ positive observations at the three considered stations, respectively.  Figure~\ref{fig:scatter_precip} shows pairwise scatterplots of the daily precipitation totals (mm) scaled by their empirical standard deviations for each station pair, highlighting the dependence across the full intensity range. %shows the pairwise scatterplots of the scaled (by their empirical standard deviations) daily precipitations for the three pairs of stations. These plots show the asymptotic dependence between the precipitations of these stations. 

\begin{table}[t!]
\begin{center}
\caption{Station metadata: IDs, names, coordinates and elevations for the three North Brabant sites.}
\medskip
\begin{tabular}{c c c c c}
\hline
ID & Station     & Latitude (${}^\circ$N) & Longitude (${}^\circ$E) & Elev.\ (m) \\
\hline
2357  & Ammerzoden  & 51.75         & 5.2167          & 2          \\
2417  & Giersbergen & 51.65         & 5.1500          & 6          \\
2556  & Zaltbommel  & 51.80         & 5.2667          & 2          \\
\hline
\end{tabular}
\label{tab:station_info}
\end{center}
\end{table}

%\subsection{Model fitting and diagnostics}

% Prior to inference, all positive‐rainfall series were first transformed via the, variance‐stabilizing function, %$\text{signed\_log}(x) = \bigl[\operatorname{sign}(x)\ln(1 + |x|)\bigr] - 1$, 
% to mitigate skewness and improve neural‐estimator performance.  
We fit the bivariate version of our eGPD~\eqref{eq:meGPD} to each pair of stations using our trained NPE under the
same settings given in Section~\ref{sec:simulation}.  Table~\ref{tab:precip_est} reports the resulting point estimates and 95\% credible intervals for \(\kappa\), \(\sigma\), \(\xi\), \(\theta_{\bm L}\), \(\theta_{\bm U}\) and \(\theta_\omega\).  Notably, the lower‐tail parameter \(\theta_{\bm L}\), and the upper‐tail parameter \(\theta_{\bm U}\) are largest for the Ammerzoden--Zaltbommel pair, reflecting its stronger dependence at small and large rainfall amounts, while the upper‐tail is moderately heavy (as reflected by \(\xi\)) across all three pairs, in line with typical extreme rainfall behavior.  

\begin{table}[t!]
\centering\small
\caption{Neural posterior median estimates (bold) and 95\% posterior credible intervals for the multivariate eGPD parameters at each station pair.}
\begin{tabular}{lccc}
\hline
Parameter & Ammerzoden--Giersbergen & Ammerzoden--Zaltbommel & Giersbergen--Zaltbommel \\
\hline
\(\kappa\)   & \(\mathbf{1.196}\,(1.097,\,1.304)\) & \(\mathbf{1.116}\,(1.032,\,1.215)\) & \(\mathbf{1.268}\,(1.161,\,1.407)\) \\
\(\sigma\)   & \(\mathbf{1.313}\,(1.157,\,1.484)\) & \(\mathbf{1.381}\,(1.226,\,1.553)\) & \(\mathbf{1.214}\,(1.068,\,1.380)\) \\
\(\xi\)      & \(\mathbf{0.216}\,(0.157,\,0.281)\) & \(\mathbf{0.193}\,(0.136,\,0.259)\) & \(\mathbf{0.242}\,(0.176,\,0.307)\) \\
\(\theta_{\bm L}\) & \(\mathbf{3.955}\,(3.476,\,4.439)\) & \(\mathbf{4.082}\,(3.488,\,4.752)\) & \(\mathbf{3.593}\,(3.190,\,4.048)\) \\
\(\theta_{\bm U}\) & \(\mathbf{18.335}\,(16.658,\,20.336)\) & \(\mathbf{21.043}\,(19.358,\,22.980)\) & \(\mathbf{14.999}\,(13.371,\,16.881)\) \\
\(\theta_\omega\) & \(\mathbf{0.421}\,(0.263,\,0.492)\) & \(\mathbf{0.237}\,(0.083,\,0.401)\) & \(\mathbf{0.485}\,(0.417,\,0.504)\) \\
\hline
\end{tabular}
\label{tab:precip_est}
\end{table}

\begin{figure}[t!]
  \hspace{20pt}
  \includegraphics[width=1\linewidth]
  {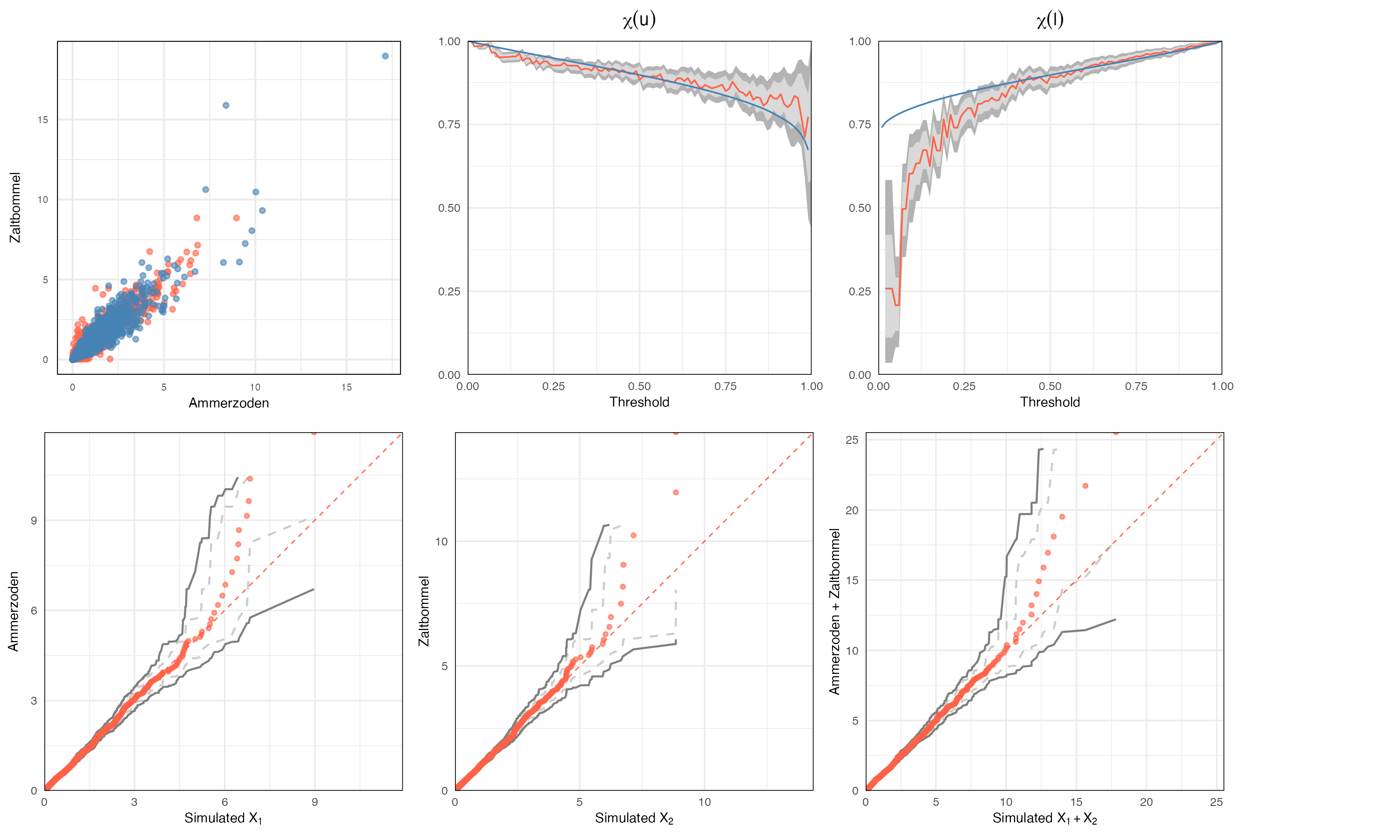}
  \caption{\small Diagnostics for the Ammerzoden--Zaltbommel fit. Top: Real (red) and simulated (blue) scatterplot based on the estimated parameters (left), and the $\chi$-dependence measures for the upper (middle) and lower (right) tails, computed based on observed (red) and simulated (blue) data, with 95\% bootstrap overall error bands (dark gray) and pointwise error bands (light grey). Bottom: QQ plots of observed versus simulated data for each margin (left and middle) and for the joint distribution summarized by the sum across stations (right). The 95\% overall error (dark gray line), and the pointwise confidence bands (dashed light gray) were obtained by bootstrapping. Recall that the data have been rescaled by their empirical standard deviations.}
  \label{fig:precip_diag}
\end{figure}

\begin{figure}[t!]
  \centering
  \includegraphics[width=0.9\linewidth]{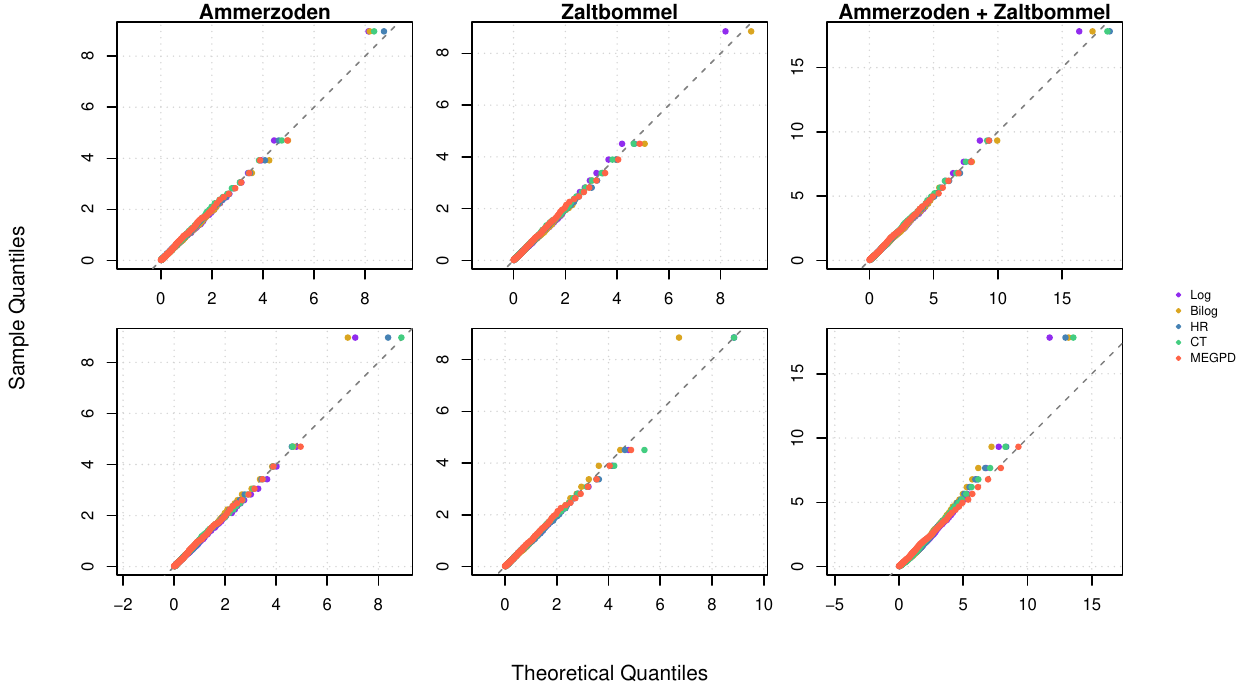}
  \caption{\small QQ-plots for marginal (left and middle) and joint (right; summarized by the sum across stations) distributions, comparing classical bivariate GP models fitted to data from the upper tail (top) or lower tail (bottom), versus our multivariate eGPD model fitted to the entire dataset, for the Ammerzoden--Zaltbommel pair. Thresholds are selected as $2.83$ mm (Ammerzoden) and $2.81$ mm (Zaltbommel) for the upper tail, and $0.082$ mm (Ammerzoden) and $0.077$ mm (Zaltbommel) for the lower tail (after rescaling by the empirical standard deviation). For readability, the QQ-plots are displayed after transforming data from each tail to a common exponential scale (so, for the lower tail, small values are to the right of each panel). Classical bivariate GP fits include the logistic (``Log''; purple), bilogistic (``Bilog''; yellow), H\"usler–Reiss (``HR''; blue), and Coles–Tawn (``CT''; green) models, along with our multivariate eGPD model (``MEGPD''; orange).}
  \label{fig:precip_compare}
\end{figure}

Diagnostic plots for the Ammerzoden--Zaltbommel fit are shown in Figure~\ref{fig:precip_diag}; results for the other two pairs appear in \ref{sec:S-diagnostics}. In the scaled scatterplot, observed points (red) and model simulations (blue) coincide closely across both the bulk and the tails, indicating that the dependence patterns are well captured. To assess the extremal dependence behavior, that is, whether large rainfall amounts at separate stations tend to occur simultaneously, we plot the $\chi$‐measure \citep{coles1999dependence} for the true (red) and the model-based (blue) joint upper-tail exceedances (and symmetrically for the joint lower-tail), together with the 95\% overall error envelopes and pointwise bootstrap confidence bands. These curves show that the model accurately reproduces the asymptotic dependence in the upper tails, while lower-tail estimates exhibit some lack-of-fit at small levels. We note that, similar to the issues pointed out by \citet{naveau2016modeling}, there are very few distinct rainfall values below the 20\% quantile (due to the limited precision of the measuring instrument), which reduces the information available for the lower tail and causes some model misspecification due to the discrete nature of the data; this likely explains the slightly larger bias observed in the lower-tail $\chi$-measure at small levels. Nevertheless, quantile-quantile (QQ)‐plots of the simulated versus observed marginal and joint distributions, together with 95\% overall error envelopes and pointwise bootstrap confidence bands, demonstrate reliable fits and extrapolation to extreme quantiles. Overall, our proposed model thus provides satisfactory marginal and dependence fits from low to high quantiles. %We acknowledge that assuming identical marginal distributions for all stations can be restrictive; introducing station‐specific scale parameters $sigma_i$ is a natural extension that could further improve model flexibility and fit.

Finally, to benchmark our modeling framework against classical extreme-value approaches, Figure~\ref{fig:precip_compare} displays marginal and joint QQ-plots for the Ammerzoden--Zaltbommel pair where we overlay our multivariate eGPD fit (based on the entire dataset) 
%, focusing only on the upper and the low joint tale (after fitted to the entire data set),
with four standard bivariate generalized Pareto (GP) models---namely, logistic (``Log''; \citealp{gumbel1960distributions}), bilogistic (``Bilog''; \citealp{smith1990extreme}), H\"usler–Reiss (``HR''; \citealp{husler1989maxima}) and Coles--Tawn (``CT''; \citealp{coles1991modelling})---fitted either to data from the upper tail (first row) or data from the lower tail (bottom row). Precisely, each classical extreme-value model is fitted separately to the lowest $10\%$ of fall–winter rainfall totals ($\leq$0.082 mm at Ammerzoden, $\leq$0.077 mm at Zaltbommel, after rescaling) and to the highest $5\%$ of totals ($\geq$2.83 mm at Ammerzoden, $\geq$2.81 mm at Zaltbommel, after rescaling). The joint and marginal fits (displayed after transforming the data to a common exponential scale, for readability) appear to capture joint high and low rainfall extremes effectively. The close agreement between the bivariate GP models and our multivariate eGPD suggests that our approach has the same capacity at representing the high and low rainfall data as the classical bivariate GP model, with the added advantage of modeling moderate rainfall effectively. In other words, our model provides an accurate description of the full range of the rainfall data in this example. 
% Our multivariate eGPD delivers a seamless fit across the entire range of rainfall intensities, from moderate through extremes.
%This unified bulk–tail performance highlights the advantage of our threshold‐free formulation.

%%%%%%%%%%%%%%%%%%%%%%%%%%%%%%%%%
%%%%%%%%%%%%%%%%%%%%%%%%%%%%%%%%%
\section{Conclusion}\label{sec:conclusions}
In this work, we analyze high and low extremes in a multivariate context, while completely bypassing the complex threshold selection step usually required in extreme-value analyses. 
 We achieve this by extending the univariate extended generalized Pareto distribution widely used in environmental modeling. 
% Our model can be viewed as an extension of the univariate eGPD class widely used in environmental modeling. 
Our proposed modeling framework offers great flexibility, as practitioners can freely choose parametric distributions for the nonnegative random vectors ${\bm L}$ and ${\bm U}$ in \eqref{eq:meGPD} (on the simplex), which characterize dependence in the lower and upper tails. 
% Still, for simulation-based inference purposes, the parameter space should not be too large and random draws from $R$, ${\bm L}$ and ${\bm U}$ should be easy to obtain. 
% Under these conditions, neural inference approaches make the estimation computationally feasible and fast, and we have illustrated through simulations how these amortized estimation methods can efficiently approximate the entire posterior distribution of model parameters or point summaries thereof.
 While parameter estimation is notoriously challenging with classical models for multivariate extremes, our model construction  allows for straightforward and fast simulation, which enables simulation-based inference methods that avoid likelihood-function computations. In particular, our simulation experiments show that neural networks can efficiently approximate the entire posterior distribution of model parameters or point summaries thereof. 
 
Despite its advantages, our modeling framework also has certain limitations. A first limitation of our model is that the vectors $\bm L$ and $\bm U$, while primarily controlling the dependence structure, also affect the marginal distributions. This may be particularly problematic when a permutation-asymmetric dependence structure must be specified, as choosing non-identically distributed components $L_j$ and/or $U_j$ would also (differently) impact the marginal scale of each variable $Y_j$. In our application, this was not a problem because we scaled the data by their empirical standard deviation and fitted a model with permutation-symmetric $\bm L$ and $\bm U$. If this is a problem in future applications, it would be easy to mitigate it by extending the multivariate eGPD model in \eqref{eq:meGPD} to allow for different marginal scale parameters. This can be achieved by defining the random vector $\bm Z=(Z_1,\ldots,Z_d)^\top$ as 
%\begin{equation}
%\label{eq:meGPD2}
$\bm Z=\bm\sigma\bm Y$, 
%\end{equation}
where  $\bm Y$ is defined as in \eqref{eq:meGPD} but setting $\sigma=1$, and where  $\bm\sigma=(\sigma_1,\ldots,\sigma_d)^\top$ with $\sigma_j>0$, $j=1,\ldots,d$, comprises component-specific scale parameters. In case the margins of $\bm L$ or $\bm U$ are not identical, this discrepancy will be absorbed by the scale parameters $\sigma_j$. 

Another related limitation is that, even with the above modification, all components $Z_j$ of $\bm Z$ (or $Y_j$ in \eqref{eq:meGPD}) of our multivariate eGPD model share the same lower tail shape parameter $\kappa>0$ and upper tail shape parameter $\xi>0$. While one might expect natural processes observed at different close-by locations to have similar tail behaviors in general, empirical evidence shows that they may vary over large or complex geographical domains. %(e.g., precipitation at far-away locations, or at different altitudes). 
If that is an issue, another %possible 
model extension allowing for component-specific scale and shape parameters can be obtained by replacing the random variable $R$ in \eqref{eq:meGPD} with a random vector $\bm{R}=(R_1,\ldots,R_d)^\top$ %defined by setting 
where $R_j=F_{R_j}^{-1}(U)$ and $F_{R_j}$ is the univariate eGPD cumulative distribution function with parameters $\kappa_j,\sigma_j,\xi_j>0$ and $U\sim{\rm Unif}(0,1)$ a uniform random variable. In this way, all components of $\bm R$ are comonotonic (i.e., perfectly positively dependent) as in the original model, but they have their own specific marginal eGPD parameters. The precise theoretical study of the joint lower/upper tails in this framework is left for future research. 

A final limitation of our proposed model is that it does not cover the case of asymptotic independence, which is often found as a plausible assumption with environmental data \citep{huser2022advances,huser2025time}. Further research is needed in this direction to develop more flexible tail models \cite[but see, e.g.,][]{gong2022asymmetric,Andre.etal:2025}.

We conclude by noting that 
%, in general, 
no unique solution exists to extend the univariate eGPD to the multivariate context, and other options should thus be explored. 
%a univariate distribution to the multivariate context. This statement also holds true for the eGPD, and other options should thus be explored. 
For example, the ongoing work of \cite{GaetanNaveau25} proposes a different type of multivariate eGPD based on a logistic-heteroscedastic model. Their model is designed so that the radial and angular components remain dependent at non-extreme levels. Another important difference between our current work and that of \cite{GaetanNaveau25} is how inference is performed. While our approach relies on an additive structure with dynamic weighting (recall \eqref{eq:meGPD}), which makes the likelihood function intractable, the heteroscedastic model of \citet{GaetanNaveau25} is amenable to likelihood-based inference when using a Gaussian vector in its construction. 
% Overall, although both approaches stem from the same univariate eGPD seed, they lead to different multivariate models, and we advise practitioners, whenever possible, to fit these two different models. Depending on the problem at hand, one model might do better than the other.
Overall, although both approaches stem from the same univariate eGPD seed, they lead to different multivariate models, each with their pros and cons.

\section*{Acknowledgments} 

Noura Alotaibi, Matthew Sainsbury-Dale, and Rapha\"el Huser were supported by King Abdullah University of Science and Technology (KAUST) under Award No. RFS-OFP2023-5550. Part of Philippe Naveau's research work was supported by the French Agence Nationale de la Recherche: EXSTA, the PEPR TRACCS (PC4 EXTENDING, ANR-22-EXTR-0005), the PEPR IRIMONT (France 2030 ANR-22-EXIR-0003) SHARE PEPR Maths-Vives (ANR-24-EXMA-0008) and from the Geolearning research chair, a joint initiative of Mines Paris and the French National Institute for Agricultural Research (INRAE).

%\appendix
%\section{Example Appendix Section}
%\label{app1}


%\begingroup
\setstretch{0.75}
\bibliographystyle{elsarticle-harv} 
\bibliography{Bibliography}

\begin{thebibliography}{99}
\expandafter\ifx\csname natexlab\endcsname\relax\def\natexlab#1{#1}\fi
\providecommand{\url}[1]{\texttt{#1}}
\providecommand{\href}[2]{#2}
\providecommand{\path}[1]{#1}
\providecommand{\DOIprefix}{doi:}
\providecommand{\ArXivprefix}{arXiv:}
\providecommand{\URLprefix}{URL: }
\providecommand{\Pubmedprefix}{pmid:}
\providecommand{\doi}[1]{\href{http://dx.doi.org/#1}{\path{#1}}}
\providecommand{\Pubmed}[1]{\href{pmid:#1}{\path{#1}}}
\providecommand{\bibinfo}[2]{#2}
\ifx\xfnm\relax \def\xfnm[#1]{\unskip,\space#1}\fi
%Type = Article
\bibitem[{Alouini et~al.(2026)Alouini, Belzile and Davison}]{Alouini2026}
\bibinfo{author}{Alouini, S.}, \bibinfo{author}{Belzile, L.R.},
  \bibinfo{author}{Davison, A.}, \bibinfo{year}{2026}.
\newblock \bibinfo{title}{Choosing the threshold in extreme value analysis}.
\newblock \bibinfo{journal}{Brazilian Journal of Probability and Statistics}
  \bibinfo{note}{To appear}.
%Type = Unpublished
\bibitem[{Andr\'e et~al.(2025)Andr\'e, Wadsworth and Huser}]{Andre.etal:2025}
\bibinfo{author}{Andr\'e, L.}, \bibinfo{author}{Wadsworth, J.L.},
  \bibinfo{author}{Huser, R.}, \bibinfo{year}{2025}.
\newblock \bibinfo{title}{Neural {Bayes} inference for complex bivariate
  extremal dependence models}.
\newblock \bibinfo{note}{ArXiv preprint arXiv:2503.23156}.
%Type = Misc
\bibitem[{Andr\'e and Tawn(2025)}]{andre2025gaussianmixturecopulasflexible}
\bibinfo{author}{Andr\'e, L.M.}, \bibinfo{author}{Tawn, J.A.},
  \bibinfo{year}{2025}.
\newblock \bibinfo{title}{Gaussian mixture copulas for flexible dependence
  modelling in the body and tails of joint distributions}.
\newblock \bibinfo{note}{ArXiv preprint arXiv:2503.06255}.
%Type = Article
\bibitem[{André et~al.(2024)André, Wadsworth and O’Hagan}]{Andr__2024}
\bibinfo{author}{André, L.}, \bibinfo{author}{Wadsworth, J.},
  \bibinfo{author}{O’Hagan, A.}, \bibinfo{year}{2024}.
\newblock \bibinfo{title}{Joint modelling of the body and tail of bivariate
  data}.
\newblock \bibinfo{journal}{Computational Statistics \& Data Analysis}
  \bibinfo{volume}{189}, \bibinfo{pages}{107841}.
%Type = Inproceedings
\bibitem[{Ardizzone et~al.(2019a)Ardizzone, Kruse, Rother and
  Köthe}]{Ardizzone_2018}
\bibinfo{author}{Ardizzone, L.}, \bibinfo{author}{Kruse, J.},
  \bibinfo{author}{Rother, C.}, \bibinfo{author}{Köthe, U.},
  \bibinfo{year}{2019}a.
\newblock \bibinfo{title}{Analyzing inverse problems with invertible neural
  networks}, in: \bibinfo{booktitle}{Proceedings of the 7th International
  Conference on Learning Representations (ICLR 2019)}.
%Type = Unpublished
\bibitem[{Ardizzone et~al.(2019b)Ardizzone, Lüth, Kruse, Rother and
  Köthe}]{Ardizzone_2019_conditional_normalising_flows}
\bibinfo{author}{Ardizzone, L.}, \bibinfo{author}{Lüth, C.},
  \bibinfo{author}{Kruse, J.}, \bibinfo{author}{Rother, C.},
  \bibinfo{author}{Köthe, U.}, \bibinfo{year}{2019}b.
\newblock \bibinfo{title}{Guided image generation with conditional invertible
  neural networks}.
\newblock \bibinfo{note}{ArXiv preprint arXiv:1907.02392}.
%Type = Article
\bibitem[{Bacro et~al.(2024)Bacro, Gaetan, Opitz and
  Toulemonde}]{bacro_multivariate_2024}
\bibinfo{author}{Bacro, J.N.}, \bibinfo{author}{Gaetan, C.},
  \bibinfo{author}{Opitz, T.}, \bibinfo{author}{Toulemonde, G.},
  \bibinfo{year}{2024}.
\newblock \bibinfo{title}{Multivariate peaks-over-threshold with latent
  variable representations of generalized {Pareto} vectors}.
\newblock \bibinfo{journal}{Extremes} \bibinfo{volume}{28},
  \bibinfo{pages}{273--300}.
%Type = Book
\bibitem[{Beirlant et~al.(2004)Beirlant, Goegebeur, Teugels and
  Segers}]{beirlant:goegebeur:teugels:segers:2004}
\bibinfo{author}{Beirlant, J.}, \bibinfo{author}{Goegebeur, Y.},
  \bibinfo{author}{Teugels, J.}, \bibinfo{author}{Segers, J.},
  \bibinfo{year}{2004}.
\newblock \bibinfo{title}{Statistics of Extremes}.
\newblock \bibinfo{publisher}{John Wiley \& Sons Ltd.},
  \bibinfo{address}{Chichester}.
%Type = Article
\bibitem[{Bortot and Gaetan(2024)}]{Bortot.Gaetan:2022}
\bibinfo{author}{Bortot, P.}, \bibinfo{author}{Gaetan, C.},
  \bibinfo{year}{2024}.
\newblock \bibinfo{title}{A model for space-time threshold exceedances with an
  application to extreme rainfall}.
\newblock \bibinfo{journal}{Statistical Modelling} \bibinfo{volume}{24},
  \bibinfo{pages}{169--193}.
%Type = Article
\bibitem[{Breiman(1965)}]{Breiman:1965}
\bibinfo{author}{Breiman, L.}, \bibinfo{year}{1965}.
\newblock \bibinfo{title}{On some limit theorems similar to the arc-sin law}.
\newblock \bibinfo{journal}{Theory of Probability \& Its Applications}
  \bibinfo{volume}{10}, \bibinfo{pages}{323--331}.
%Type = Article
\bibitem[{Brunner et~al.(2020)Brunner, Papalexiou, Clark and
  Gilleland}]{Brunner.etal:2020}
\bibinfo{author}{Brunner, M.I.}, \bibinfo{author}{Papalexiou, S.},
  \bibinfo{author}{Clark, M.P.}, \bibinfo{author}{Gilleland, E.},
  \bibinfo{year}{2020}.
\newblock \bibinfo{title}{How probable is widespread flooding in the {United}
  {States}?}
\newblock \bibinfo{journal}{Water Resources Research} \bibinfo{volume}{56},
  \bibinfo{pages}{e2020WR028096}.
%Type = Article
\bibitem[{Carreau and Bengio(2008)}]{Carreau2008}
\bibinfo{author}{Carreau, J.}, \bibinfo{author}{Bengio, Y.},
  \bibinfo{year}{2008}.
\newblock \bibinfo{title}{{A hybrid Pareto model for asymmetric fat-tailed
  data: the univariate case}}.
\newblock \bibinfo{journal}{Extremes} \bibinfo{volume}{12},
  \bibinfo{pages}{53--76}.
%Type = Article
\bibitem[{Carreau and Bengio(2009)}]{Carreau2009}
\bibinfo{author}{Carreau, J.}, \bibinfo{author}{Bengio, Y.},
  \bibinfo{year}{2009}.
\newblock \bibinfo{title}{{A hybrid {P}areto mixture for conditional asymmetric
  fat-tailed distributions}}.
\newblock \bibinfo{journal}{IEEE Transactions on Neural Networks}
  \bibinfo{volume}{20}, \bibinfo{pages}{1087--1101}.
%Type = Book
\bibitem[{de~Carvalho et~al.(2025)de~Carvalho, Huser, Naveau and
  Reich}]{Carvalho25}
\bibinfo{author}{de~Carvalho, M.}, \bibinfo{author}{Huser, R.},
  \bibinfo{author}{Naveau, P.}, \bibinfo{author}{Reich, B.},
  \bibinfo{year}{2025}.
\newblock \bibinfo{title}{Handbook on Statistics of Extremes}.
\newblock \bibinfo{publisher}{Chapman~\& Hall / CRC}.
%Type = Book
\bibitem[{Casella and Berger(2001)}]{Casella_2001}
\bibinfo{author}{Casella, G.}, \bibinfo{author}{Berger, R.},
  \bibinfo{year}{2001}.
\newblock \bibinfo{title}{Statistical Inference}.
\newblock \bibinfo{edition}{Second} ed., \bibinfo{publisher}{Duxbury},
  \bibinfo{address}{Belmont, CA}.
%Type = Inproceedings
\bibitem[{Chan et~al.(2018)Chan, Perrone, Spence and \emph{et~al.}}]{Chan_2018}
\bibinfo{author}{Chan, J.}, \bibinfo{author}{Perrone, V.},
  \bibinfo{author}{Spence, J.}, \bibinfo{author}{\emph{et~al.}},
  \bibinfo{year}{2018}.
\newblock \bibinfo{title}{A likelihood-free inference framework for population
  genetic data using exchangeable neural networks}, in:
  \bibinfo{editor}{Bengio, S.}, \bibinfo{editor}{Wallach, H.},
  \bibinfo{editor}{Larochelle, H.}, \bibinfo{editor}{Grauman, K.},
  \bibinfo{editor}{Cesa-Bianchi, N.}, \bibinfo{editor}{Garnett, R.} (Eds.),
  \bibinfo{booktitle}{Advances in Neural Information Processing Systems}, pp.
  \bibinfo{pages}{8594--8605}.
%Type = Article
\bibitem[{Cisneros et~al.(2024)Cisneros, Richards, Dahal, Lombardo and
  Huser}]{cisneros2024deep}
\bibinfo{author}{Cisneros, D.}, \bibinfo{author}{Richards, J.},
  \bibinfo{author}{Dahal, A.}, \bibinfo{author}{Lombardo, L.},
  \bibinfo{author}{Huser, R.}, \bibinfo{year}{2024}.
\newblock \bibinfo{title}{Deep graphical regression for jointly moderate and
  extreme {Australian} wildfires}.
\newblock \bibinfo{journal}{Spatial Statistics} \bibinfo{volume}{59},
  \bibinfo{pages}{100811}.
%Type = Book
\bibitem[{Coles(2001)}]{coles_introduction_2001}
\bibinfo{author}{Coles, S.}, \bibinfo{year}{2001}.
\newblock \bibinfo{title}{An {Introduction} to {Statistical} {Modeling} of
  {Extreme} {Values}}.
\newblock \bibinfo{publisher}{Springer}, \bibinfo{address}{New York}.
%Type = Article
\bibitem[{Coles et~al.(1999)Coles, Heffernan and Tawn}]{coles1999dependence}
\bibinfo{author}{Coles, S.}, \bibinfo{author}{Heffernan, J.},
  \bibinfo{author}{Tawn, J.}, \bibinfo{year}{1999}.
\newblock \bibinfo{title}{Dependence measures for extreme value analyses}.
\newblock \bibinfo{journal}{Extremes} \bibinfo{volume}{2},
  \bibinfo{pages}{339--365}.
%Type = Article
\bibitem[{Coles and Tawn(1991)}]{coles1991modelling}
\bibinfo{author}{Coles, S.G.}, \bibinfo{author}{Tawn, J.A.},
  \bibinfo{year}{1991}.
\newblock \bibinfo{title}{Modelling extreme multivariate events}.
\newblock \bibinfo{journal}{Journal of the Royal Statistical Society: Series B
  (Statistical Methodology)} \bibinfo{volume}{53}, \bibinfo{pages}{377--392}.
%Type = Unpublished
\bibitem[{Coloma and
  Kleiber(2025)}]{coloma2025fastlikelihoodfreeparameterestimation}
\bibinfo{author}{Coloma, N.}, \bibinfo{author}{Kleiber, W.},
  \bibinfo{year}{2025}.
\newblock \bibinfo{title}{Fast likelihood-free parameter estimation for
  {L}\'evy processes}.
\newblock \bibinfo{note}{ArXiv preprint arXiv.2501.01639}.
%Type = Article
\bibitem[{Davison and Huser(2015)}]{Davison.Huser:2015}
\bibinfo{author}{Davison, A.C.}, \bibinfo{author}{Huser, R.},
  \bibinfo{year}{2015}.
\newblock \bibinfo{title}{{Statistics of Extremes}}.
\newblock \bibinfo{journal}{{Annual Review of Statistics and its Application}}
  \bibinfo{volume}{2}, \bibinfo{pages}{203--235}.
%Type = Article
\bibitem[{Davison and Smith(1990)}]{Davison.Smith:1990}
\bibinfo{author}{Davison, A.C.}, \bibinfo{author}{Smith, R.L.},
  \bibinfo{year}{1990}.
\newblock \bibinfo{title}{{Models for exceedances over high thresholds (with
  discussion)}}.
\newblock \bibinfo{journal}{{Journal of the Royal Statistical Society: Series B
  (Statistical Methodology)}} \bibinfo{volume}{52}, \bibinfo{pages}{393--442}.
%Type = Inproceedings
\bibitem[{Dinh et~al.(2016)Dinh, Sohl-Dickstein and Bengio}]{Dinh_2016}
\bibinfo{author}{Dinh, L.}, \bibinfo{author}{Sohl-Dickstein, J.},
  \bibinfo{author}{Bengio, S.}, \bibinfo{year}{2016}.
\newblock \bibinfo{title}{Density estimation using {Real NVP}}, in:
  \bibinfo{booktitle}{Proceedings of the 4th International Conference on
  Learning Representations (ICLR 2016)}.
%Type = Book
\bibitem[{Embrechts et~al.(2013)Embrechts, Kl{\"u}ppelberg and
  Mikosch}]{embrechts2013modelling}
\bibinfo{author}{Embrechts, P.}, \bibinfo{author}{Kl{\"u}ppelberg, C.},
  \bibinfo{author}{Mikosch, T.}, \bibinfo{year}{2013}.
\newblock \bibinfo{title}{Modelling Extremal Events: for Insurance and
  Finance}. volume~\bibinfo{volume}{33}.
\newblock \bibinfo{publisher}{Springer Berlin Heidelberg}.
%Type = Article
\bibitem[{Engelke et~al.(2019)Engelke, Opitz and Wadsworth}]{engelke2019}
\bibinfo{author}{Engelke, S.}, \bibinfo{author}{Opitz, T.},
  \bibinfo{author}{Wadsworth, J.}, \bibinfo{year}{2019}.
\newblock \bibinfo{title}{Extremal dependence of random scale constructions}.
\newblock \bibinfo{journal}{Extremes} \bibinfo{volume}{22},
  \bibinfo{pages}{623--666}.
%Type = Article
\bibitem[{Evin et~al.(2018)Evin, Favre and Hingray}]{evin2018}
\bibinfo{author}{Evin, G.}, \bibinfo{author}{Favre, A.C.},
  \bibinfo{author}{Hingray, B.}, \bibinfo{year}{2018}.
\newblock \bibinfo{title}{Stochastic generation of multi-site daily
  precipitation focusing on extreme events}.
\newblock \bibinfo{journal}{Hydrology and Earth System Sciences}
  \bibinfo{volume}{22}, \bibinfo{pages}{655--672}.
%Type = Article
\bibitem[{Fischer et~al.(2025)Fischer, Bador, Huser, Kendon, Robinson and
  Sippel}]{fischer2025}
\bibinfo{author}{Fischer, E.M.}, \bibinfo{author}{Bador, M.},
  \bibinfo{author}{Huser, R.}, \bibinfo{author}{Kendon, E.J.},
  \bibinfo{author}{Robinson, A.}, \bibinfo{author}{Sippel, S.},
  \bibinfo{year}{2025}.
\newblock \bibinfo{title}{Record-breaking extremes in a warming climate}.
\newblock \bibinfo{journal}{Nature Reviews Earth \& Environment}
  \bibinfo{volume}{6}, \bibinfo{pages}{456--470}.
%Type = Article
\bibitem[{de~Fondeville and Davison(2018)}]{deFondevilleDavison18}
\bibinfo{author}{de~Fondeville, R.}, \bibinfo{author}{Davison, A.C.},
  \bibinfo{year}{2018}.
\newblock \bibinfo{title}{High-dimensional peaks-over-threshold inference}.
\newblock \bibinfo{journal}{Biometrika} \bibinfo{volume}{105},
  \bibinfo{pages}{575--592}.
%Type = Article
\bibitem[{Foug{\`e}res and Mercadier(2012)}]{Fougeres:Mercadier:2012}
\bibinfo{author}{Foug{\`e}res, A.L.}, \bibinfo{author}{Mercadier, C.},
  \bibinfo{year}{2012}.
\newblock \bibinfo{title}{Risk measures and multivariate extensions of
  {B}reiman's theorem}.
\newblock \bibinfo{journal}{Journal of Applied Probability}
  \bibinfo{volume}{49}, \bibinfo{pages}{364--384}.
%Type = Article
\bibitem[{Frigessi et~al.(2002)Frigessi, Haug and Rue}]{Frigessi2002}
\bibinfo{author}{Frigessi, A.}, \bibinfo{author}{Haug, O.},
  \bibinfo{author}{Rue, H.}, \bibinfo{year}{2002}.
\newblock \bibinfo{title}{{A dynamic mixture model for unsupervised tail
  estimation without threshold selection}}.
\newblock \bibinfo{journal}{Extremes} \bibinfo{volume}{5},
  \bibinfo{pages}{219--235}.
%Type = Misc
\bibitem[{Gaetan and Naveau(2025)}]{GaetanNaveau25}
\bibinfo{author}{Gaetan, C.}, \bibinfo{author}{Naveau, P.},
  \bibinfo{year}{2025}.
\newblock \bibinfo{title}{Multivariate distributional modeling of low,
  moderate, and large intensities without threshold selection steps}.
\newblock \bibinfo{note}{In preparation}.
%Type = Article
\bibitem[{Gamet and Jalbert(2022)}]{Gamet22}
\bibinfo{author}{Gamet, P.}, \bibinfo{author}{Jalbert, J.},
  \bibinfo{year}{2022}.
\newblock \bibinfo{title}{A flexible extended generalized {P}areto distribution
  for tail estimation}.
\newblock \bibinfo{journal}{Environmetrics} \bibinfo{volume}{33},
  \bibinfo{pages}{e2744}.
%Type = Article
\bibitem[{Gong and Huser(2022)}]{gong2022asymmetric}
\bibinfo{author}{Gong, Y.}, \bibinfo{author}{Huser, R.}, \bibinfo{year}{2022}.
\newblock \bibinfo{title}{Asymmetric tail dependence modeling, with application
  to cryptocurrency market data}.
\newblock \bibinfo{journal}{The Annals of Applied Statistics}
  \bibinfo{volume}{16}, \bibinfo{pages}{1822--1847}.
%Type = Article
\bibitem[{Gumbel(1960)}]{gumbel1960distributions}
\bibinfo{author}{Gumbel, E.J.}, \bibinfo{year}{1960}.
\newblock \bibinfo{title}{Distributions des valeurs extremes en plusiers
  dimensions}.
\newblock \bibinfo{journal}{Annales de l'ISUP} \bibinfo{volume}{9},
  \bibinfo{pages}{171--173}.
%Type = Book
\bibitem[{de~Haan and Ferreira(2006)}]{dehaan:ferreira:2006}
\bibinfo{author}{de~Haan, L.}, \bibinfo{author}{Ferreira, A.},
  \bibinfo{year}{2006}.
\newblock \bibinfo{title}{Extreme Value Theory: An Introduction}.
\newblock \bibinfo{publisher}{Springer-Verlag}, \bibinfo{address}{New York}.
%Type = Article
\bibitem[{Haruna et~al.(2023)Haruna, Blanchet and Favre}]{haruna2023modeling}
\bibinfo{author}{Haruna, A.}, \bibinfo{author}{Blanchet, J.},
  \bibinfo{author}{Favre, A.C.}, \bibinfo{year}{2023}.
\newblock \bibinfo{title}{Modeling intensity-duration-frequency curves for the
  whole range of non-zero precipitation: a comparison of models}.
\newblock \bibinfo{journal}{Water Resources Research} \bibinfo{volume}{59},
  \bibinfo{pages}{e2022WR033362}.
%Type = Article
\bibitem[{Haruna et~al.(2024)Haruna, Blanchet and Favre}]{Haruna24}
\bibinfo{author}{Haruna, A.}, \bibinfo{author}{Blanchet, J.},
  \bibinfo{author}{Favre, A.C.}, \bibinfo{year}{2024}.
\newblock \bibinfo{title}{Estimation of intensity-duration-area-frequency
  relationships based on the full range of non-zero precipitation from
  radar-reanalysis data.}
\newblock \bibinfo{journal}{Water Resources Research} \bibinfo{volume}{60},
  \bibinfo{pages}{e2023WR035902}.
%Type = Article
\bibitem[{Hazra et~al.(2025)Hazra, Huser and Bolin}]{hazra2024efficient}
\bibinfo{author}{Hazra, A.}, \bibinfo{author}{Huser, R.},
  \bibinfo{author}{Bolin, D.}, \bibinfo{year}{2025}.
\newblock \bibinfo{title}{Efficient modeling of spatial extremes over large
  geographical domains}.
\newblock \bibinfo{journal}{Journal of Computational and Graphical Statistics}
  \bibinfo{volume}{34}, \bibinfo{pages}{795--811}.
%Type = Article
\bibitem[{Hornik et~al.(1989)Hornik, Stinchcombe and
  White}]{Hornik_1989_FNN_universal_approximation_theorem}
\bibinfo{author}{Hornik, K.}, \bibinfo{author}{Stinchcombe, M.},
  \bibinfo{author}{White, H.}, \bibinfo{year}{1989}.
\newblock \bibinfo{title}{Multilayer feedforward networks are universal
  approximators}.
\newblock \bibinfo{journal}{Neural Networks} \bibinfo{volume}{2},
  \bibinfo{pages}{359--366}.
%Type = Article
\bibitem[{Huser et~al.(2017)Huser, Opitz and Thibaud}]{huser2017bridging}
\bibinfo{author}{Huser, R.}, \bibinfo{author}{Opitz, T.},
  \bibinfo{author}{Thibaud, E.}, \bibinfo{year}{2017}.
\newblock \bibinfo{title}{Bridging asymptotic independence and dependence in
  spatial extremes using {G}aussian scale mixtures}.
\newblock \bibinfo{journal}{Spatial Statistics} \bibinfo{volume}{21},
  \bibinfo{pages}{166--186}.
%Type = Article
\bibitem[{Huser et~al.(2025)Huser, Opitz and Wadsworth}]{huser2025time}
\bibinfo{author}{Huser, R.}, \bibinfo{author}{Opitz, T.},
  \bibinfo{author}{Wadsworth, J.L.}, \bibinfo{year}{2025}.
\newblock \bibinfo{title}{Modeling of spatial extremes in environmental data
  science: Time to move away from max-stable processes}.
\newblock \bibinfo{journal}{Environmental Data Science} \bibinfo{volume}{4},
  \bibinfo{pages}{1--16}.
%Type = Article
\bibitem[{Huser and Wadsworth(2022)}]{huser2022advances}
\bibinfo{author}{Huser, R.}, \bibinfo{author}{Wadsworth, J.L.},
  \bibinfo{year}{2022}.
\newblock \bibinfo{title}{Advances in statistical modeling of spatial
  extremes}.
\newblock \bibinfo{journal}{Wiley Interdisciplinary Reviews: Computational
  Statistics} \bibinfo{volume}{14}, \bibinfo{pages}{e1537}.
%Type = Article
\bibitem[{H{\"u}sler and Reiss(1989)}]{husler1989maxima}
\bibinfo{author}{H{\"u}sler, J.}, \bibinfo{author}{Reiss, R.D.},
  \bibinfo{year}{1989}.
\newblock \bibinfo{title}{Maxima of normal random vectors: between independence
  and complete dependence}.
\newblock \bibinfo{journal}{Statistics \& Probability Letters}
  \bibinfo{volume}{7}, \bibinfo{pages}{283--286}.
%Type = Incollection
\bibitem[{IPCC(2023)}]{ipcc2023}
\bibinfo{author}{IPCC}, \bibinfo{year}{2023}.
\newblock \bibinfo{title}{{Climate Change 2023: Synthesis Report}}, in:
  \bibinfo{editor}{Core Writing~Team, H.L.}, \bibinfo{editor}{Romero, J.}
  (Eds.), \bibinfo{booktitle}{Contribution of Working Groups I, II, and III to
  the Sixth Assessment Report of the Intergovernmental Panel on Climate
  Change}. \bibinfo{publisher}{IPCC, Geneva, Switzerland}, pp.
  \bibinfo{pages}{35--115}.
%Type = Article
\bibitem[{Kakampakou et~al.(2024)Kakampakou, Simpson and
  Wadsworth}]{kakampakou2024spatial}
\bibinfo{author}{Kakampakou, L.}, \bibinfo{author}{Simpson, E.S.},
  \bibinfo{author}{Wadsworth, J.L.}, \bibinfo{year}{2024}.
\newblock \bibinfo{title}{Spatial extremal modelling: A case study on the
  interplay between margins and dependence}.
\newblock \bibinfo{journal}{Stat} \bibinfo{volume}{13},
  \bibinfo{pages}{e70021}.
%Type = Inproceedings
\bibitem[{Kingma and Dhariwal(2018)}]{Kingma_Dhariwal_2018}
\bibinfo{author}{Kingma, D.P.}, \bibinfo{author}{Dhariwal, P.},
  \bibinfo{year}{2018}.
\newblock \bibinfo{title}{Glow: Generative flow with invertible 1x1
  convolutions}, in: \bibinfo{editor}{Bengio, S.}, \bibinfo{editor}{Wallach,
  H.}, \bibinfo{editor}{Larochelle, H.}, \bibinfo{editor}{Grauman, K.},
  \bibinfo{editor}{Cesa-Bianchi, N.}, \bibinfo{editor}{Garnett, R.} (Eds.),
  \bibinfo{booktitle}{Advances in Neural Information Processing Systems},
  \bibinfo{publisher}{Curran Associates, Inc.}. pp.
  \bibinfo{pages}{10215--10224}.
%Type = Article
\bibitem[{Klein~Tank et~al.(2002)Klein~Tank, Wijngaard, K\"{o}nnen and
  \emph{et~al.}}]{tank2002publications}
\bibinfo{author}{Klein~Tank, A.M.G.}, \bibinfo{author}{Wijngaard, J.B.},
  \bibinfo{author}{K\"{o}nnen, G.P.}, \bibinfo{author}{\emph{et~al.}},
  \bibinfo{year}{2002}.
\newblock \bibinfo{title}{Daily dataset of 20th‐century surface air
  temperature and precipitation series for the {E}uropean {C}limate
  {A}ssessment}.
\newblock \bibinfo{journal}{International Journal of Climatology}
  \bibinfo{volume}{22}, \bibinfo{pages}{1441--1453}.
%Type = Article
\bibitem[{Kobyzev et~al.(2020)Kobyzev, Prince and Brubaker}]{Kobyzev_2020}
\bibinfo{author}{Kobyzev, I.}, \bibinfo{author}{Prince, S.J.},
  \bibinfo{author}{Brubaker, M.A.}, \bibinfo{year}{2020}.
\newblock \bibinfo{title}{Normalizing flows: An introduction and review of
  current methods}.
\newblock \bibinfo{journal}{IEEE Transactions on Pattern Analysis and Machine
  Intelligence} \bibinfo{volume}{43}, \bibinfo{pages}{3964--3979}.
%Type = Article
\bibitem[{Krupskii and Huser(2022)}]{Krupskii.Huser:2021}
\bibinfo{author}{Krupskii, P.}, \bibinfo{author}{Huser, R.},
  \bibinfo{year}{2022}.
\newblock \bibinfo{title}{{Modeling spatial tail dependence with Cauchy
  convolution processes}}.
\newblock \bibinfo{journal}{Electronic Journal of Statistics}
  \bibinfo{volume}{16}, \bibinfo{pages}{6135--6174}.
%Type = Book
\bibitem[{Kulik and Soulier(2020)}]{kulik_heavy-tailed_2020}
\bibinfo{author}{Kulik, R.}, \bibinfo{author}{Soulier, P.},
  \bibinfo{year}{2020}.
\newblock \bibinfo{title}{Heavy-{Tailed} {Time} {Series}}.
\newblock \bibinfo{publisher}{Springer}, \bibinfo{address}{New York}.
%Type = Article
\bibitem[{Kullback and Leibler(1951)}]{Kullback_1951}
\bibinfo{author}{Kullback, S.}, \bibinfo{author}{Leibler, R.A.},
  \bibinfo{year}{1951}.
\newblock \bibinfo{title}{On information and sufficiency}.
\newblock \bibinfo{journal}{Annals of Mathematical Statistics}
  \bibinfo{volume}{22}, \bibinfo{pages}{79--86}.
%Type = Article
\bibitem[{{Le Gall} et~al.(2022){Le Gall}, Favre, Naveau and
  Prieur}]{Legall2022}
\bibinfo{author}{{Le Gall}, P.}, \bibinfo{author}{Favre, A.C.},
  \bibinfo{author}{Naveau, P.}, \bibinfo{author}{Prieur, C.},
  \bibinfo{year}{2022}.
\newblock \bibinfo{title}{Improved regional frequency analysis of rainfall
  data}.
\newblock \bibinfo{journal}{Weather and Climate Extremes} \bibinfo{volume}{36},
  \bibinfo{pages}{100456}.
%Type = Article
\bibitem[{Lenzi et~al.(2023)Lenzi, Bessac, Rudi and Stein}]{lenzi2023neural}
\bibinfo{author}{Lenzi, A.}, \bibinfo{author}{Bessac, J.},
  \bibinfo{author}{Rudi, J.}, \bibinfo{author}{Stein, M.L.},
  \bibinfo{year}{2023}.
\newblock \bibinfo{title}{Neural networks for parameter estimation in
  intractable models}.
\newblock \bibinfo{journal}{Computational Statistics and Data Analysis}
  \bibinfo{volume}{185}, \bibinfo{pages}{107762}.
%Type = Unpublished
\bibitem[{Maceda et~al.(2024)Maceda, Hector, Lenzi and Reich}]{Maceda_2024}
\bibinfo{author}{Maceda, E.}, \bibinfo{author}{Hector, E.C.},
  \bibinfo{author}{Lenzi, A.}, \bibinfo{author}{Reich, B.J.},
  \bibinfo{year}{2024}.
\newblock \bibinfo{title}{A variational neural {B}ayes framework for inference
  on intractable posterior distributions}.
\newblock \bibinfo{note}{ArXiv preprint arXiv.2404.10899}.
%Type = Unpublished
\bibitem[{Mackay and Jonathan(2024)}]{Mackay:Jonathan:2024}
\bibinfo{author}{Mackay, E.}, \bibinfo{author}{Jonathan, P.},
  \bibinfo{year}{2024}.
\newblock \bibinfo{title}{Modelling multivariate extremes through
  angular-radial decomposition of the density function}.
\newblock \bibinfo{note}{ArXiv preprint arXiv:2310.12711}.
%Type = Article
\bibitem[{McFadden(1989)}]{McFadden1989}
\bibinfo{author}{McFadden, D.}, \bibinfo{year}{1989}.
\newblock \bibinfo{title}{A method of simulated moments for estimation of
  discrete response models without numerical integration}.
\newblock \bibinfo{journal}{Econometrica} \bibinfo{volume}{57},
  \bibinfo{pages}{995--1026}.
%Type = Article
\bibitem[{Merz et~al.(2021)Merz, Bl{\"o}schl, Vorogushyn, Dottori, Aerts,
  Bates, Bertola, Kemter, Kreibich, Lall and Macdonald}]{Merz.etal:2021}
\bibinfo{author}{Merz, B.}, \bibinfo{author}{Bl{\"o}schl, G.},
  \bibinfo{author}{Vorogushyn, S.}, \bibinfo{author}{Dottori, F.},
  \bibinfo{author}{Aerts, J.C.J.H.}, \bibinfo{author}{Bates, P.},
  \bibinfo{author}{Bertola, M.}, \bibinfo{author}{Kemter, M.},
  \bibinfo{author}{Kreibich, H.}, \bibinfo{author}{Lall, U.},
  \bibinfo{author}{Macdonald, E.}, \bibinfo{year}{2021}.
\newblock \bibinfo{title}{Causes, impacts and patterns of disastrous river
  floods}.
\newblock \bibinfo{journal}{Nature Reviews Earth \& Environment}
  \bibinfo{volume}{2}, \bibinfo{pages}{592--609}.
%Type = Article
\bibitem[{Modrák et~al.(2025)Modrák, Moon, Kim, Bürkner, Huurre,
  Faltejsková, Gelman and Vehtari}]{Modrak_2025}
\bibinfo{author}{Modrák, M.}, \bibinfo{author}{Moon, A.H.},
  \bibinfo{author}{Kim, S.}, \bibinfo{author}{Bürkner, P.},
  \bibinfo{author}{Huurre, N.}, \bibinfo{author}{Faltejsková, K.},
  \bibinfo{author}{Gelman, A.}, \bibinfo{author}{Vehtari, A.},
  \bibinfo{year}{2025}.
\newblock \bibinfo{title}{Simulation-based calibration checking for {B}ayesian
  computation: The choice of test quantities shapes sensitivity}.
\newblock \bibinfo{journal}{Bayesian Analysis} \bibinfo{volume}{20},
  \bibinfo{pages}{461--488}.
%Type = Article
\bibitem[{Morris et~al.(2017)Morris, Reich, Thibaud and
  Cooley}]{morris2017space}
\bibinfo{author}{Morris, S.A.}, \bibinfo{author}{Reich, B.J.},
  \bibinfo{author}{Thibaud, E.}, \bibinfo{author}{Cooley, D.},
  \bibinfo{year}{2017}.
\newblock \bibinfo{title}{A space-time skew-t model for threshold exceedances}.
\newblock \bibinfo{journal}{Biometrics} \bibinfo{volume}{73},
  \bibinfo{pages}{749--758}.
%Type = Article
\bibitem[{Murphy-Barltrop et~al.(2024)Murphy-Barltrop, Mackay and
  Jonathan}]{murphy2024inference}
\bibinfo{author}{Murphy-Barltrop, C.J.R.}, \bibinfo{author}{Mackay, E.},
  \bibinfo{author}{Jonathan, P.}, \bibinfo{year}{2024}.
\newblock \bibinfo{title}{Inference for bivariate extremes via a
  semi-parametric angular-radial model}.
\newblock \bibinfo{journal}{Extremes} , \bibinfo{pages}{1--30}.
%Type = Incollection
\bibitem[{Naveau(2025)}]{Naveau25}
\bibinfo{author}{Naveau, P.}, \bibinfo{year}{2025}.
\newblock \bibinfo{title}{Jointly modeling the bulk and tails}, in:
  \bibinfo{editor}{de~Carvalho, M.}, \bibinfo{editor}{Huser, R.},
  \bibinfo{editor}{Naveau, P.}, \bibinfo{editor}{Reich, B.} (Eds.),
  \bibinfo{booktitle}{Handbook on Statistics of Extremes}.
  \bibinfo{publisher}{CRC Press}. chapter~\bibinfo{chapter}{5}.
%Type = Article
\bibitem[{Naveau et~al.(2016)Naveau, Huser, Ribereau and
  Hannart}]{naveau2016modeling}
\bibinfo{author}{Naveau, P.}, \bibinfo{author}{Huser, R.},
  \bibinfo{author}{Ribereau, P.}, \bibinfo{author}{Hannart, A.},
  \bibinfo{year}{2016}.
\newblock \bibinfo{title}{Modeling jointly low, moderate and heavy rainfall
  intensities without a threshold selection}.
\newblock \bibinfo{journal}{Water Resources Research} \bibinfo{volume}{52},
  \bibinfo{pages}{2753--2769}.
%Type = Inproceedings
\bibitem[{Papamakarios and Murray(2016)}]{Papamakarios_Murray_2016}
\bibinfo{author}{Papamakarios, G.}, \bibinfo{author}{Murray, I.},
  \bibinfo{year}{2016}.
\newblock \bibinfo{title}{Fast $\epsilon$-free inference of simulation models
  with {Bayesian} conditional density estimation}, in: \bibinfo{editor}{Lee,
  D.}, \bibinfo{editor}{Sugiyama, M.}, \bibinfo{editor}{Luxburg, U.},
  \bibinfo{editor}{Guyon, I.}, \bibinfo{editor}{Garnett, R.} (Eds.),
  \bibinfo{booktitle}{Advances in Neural Information Processing Systems},
  \bibinfo{publisher}{Curran Associates, Inc., Red Hook, NY}. pp.
  \bibinfo{pages}{1028--1036}.
%Type = Article
\bibitem[{Papamakarios et~al.(2021)Papamakarios, Nalisnick, Rezende and
  Lakshminarayanan}]{Papamakarios_2021_review}
\bibinfo{author}{Papamakarios, G.}, \bibinfo{author}{Nalisnick, E.},
  \bibinfo{author}{Rezende, D.J.}, \bibinfo{author}{Lakshminarayanan, S.M.B.},
  \bibinfo{year}{2021}.
\newblock \bibinfo{title}{Normalizing flows for probabilistic modeling and
  inference}.
\newblock \bibinfo{journal}{Journal of Machine Learning Research}
  \bibinfo{volume}{22}, \bibinfo{pages}{1--64}.
%Type = Article
\bibitem[{Papastathopoulos and Tawn(2013)}]{Papastathopoulos:Tawn:2013}
\bibinfo{author}{Papastathopoulos, I.}, \bibinfo{author}{Tawn, J.A.},
  \bibinfo{year}{2013}.
\newblock \bibinfo{title}{Extended generalised {Pareto} models for tail
  estimation}.
\newblock \bibinfo{journal}{Journal of Statistical Planning and Inference}
  \bibinfo{volume}{143}, \bibinfo{pages}{131--143}.
%Type = Article
\bibitem[{Radev et~al.(2020)Radev, Mertens, Voss, Ardizzone and
  K{\"o}the}]{radev2020bayesflow}
\bibinfo{author}{Radev, S.T.}, \bibinfo{author}{Mertens, U.K.},
  \bibinfo{author}{Voss, A.}, \bibinfo{author}{Ardizzone, L.},
  \bibinfo{author}{K{\"o}the, U.}, \bibinfo{year}{2020}.
\newblock \bibinfo{title}{{BayesFlow}: Learning complex stochastic models with
  invertible neural networks}.
\newblock \bibinfo{journal}{IEEE Transactions on Neural Networks and Learning
  Systems} \bibinfo{volume}{33}, \bibinfo{pages}{1452--1466}.
%Type = Article
\bibitem[{Rentschler et~al.(2022)Rentschler, Salhab and Jafino}]{Jun.etal:2022}
\bibinfo{author}{Rentschler, J.}, \bibinfo{author}{Salhab, M.},
  \bibinfo{author}{Jafino, B.A.}, \bibinfo{year}{2022}.
\newblock \bibinfo{title}{Flood exposure and poverty in 188 countries}.
\newblock \bibinfo{journal}{Nature Communications} \bibinfo{volume}{13},
  \bibinfo{pages}{3527}.
%Type = Book
\bibitem[{Resnick(1987)}]{Resnick1987}
\bibinfo{author}{Resnick, S.I.}, \bibinfo{year}{1987}.
\newblock \bibinfo{title}{Extreme Values, Regular Variation and Point
  Processes}.
\newblock \bibinfo{publisher}{Spinger-Verlag New York}.
%Type = Article
\bibitem[{Resnick and St\v{a}ric\v{a}(1997)}]{ResnickStarica97}
\bibinfo{author}{Resnick, S.I.}, \bibinfo{author}{St\v{a}ric\v{a}, C.},
  \bibinfo{year}{1997}.
\newblock \bibinfo{title}{Smoothing the {Hill} estimator}.
\newblock \bibinfo{journal}{Advances in Applied Probability}
  \bibinfo{volume}{29}, \bibinfo{pages}{271--293}.
%Type = Article
\bibitem[{Richards et~al.(2025)Richards, Sainsbury-Dale, Zammit-Mangion and
  Huser}]{Richards.etal:2024}
\bibinfo{author}{Richards, J.}, \bibinfo{author}{Sainsbury-Dale, M.},
  \bibinfo{author}{Zammit-Mangion, A.}, \bibinfo{author}{Huser, R.},
  \bibinfo{year}{2025}.
\newblock \bibinfo{title}{Neural {Bayes} estimators for censored inference with
  peaks-over-threshold models}.
\newblock \bibinfo{journal}{Journal of Machine Learning Research}
  \bibinfo{volume}{25}, \bibinfo{pages}{1--49}.
%Type = Article
\bibitem[{Rootz\'{e}n et~al.(2018)Rootz\'{e}n, Segers and
  Wadsworth}]{rootzen2018}
\bibinfo{author}{Rootz\'{e}n, H.}, \bibinfo{author}{Segers, J.},
  \bibinfo{author}{Wadsworth, J.L.}, \bibinfo{year}{2018}.
\newblock \bibinfo{title}{Multivariate peaks over thresholds models}.
\newblock \bibinfo{journal}{Extremes} \bibinfo{volume}{21},
  \bibinfo{pages}{115--145}.
%Type = Misc
\bibitem[{Rödder et~al.(2025)Rödder, Hentschel and
  Engelke}]{rodder2025theoretical}
\bibinfo{author}{Rödder, A.}, \bibinfo{author}{Hentschel, M.},
  \bibinfo{author}{Engelke, S.}, \bibinfo{year}{2025}.
\newblock \bibinfo{title}{Theoretical guarantees for neural estimators in
  parametric statistics}.
\newblock \bibinfo{note}{ArXiv:2506.18508}.
%Type = Manual
\bibitem[{Sainsbury-Dale(2024)}]{NeuralEstimators}
\bibinfo{author}{Sainsbury-Dale, M.}, \bibinfo{year}{2024}.
\newblock \bibinfo{title}{{NeuralEstimators}: Likelihood-Free Parameter
  Estimation using Neural Networks}.
\newblock \bibinfo{note}{R package version 0.1.2,
  \url{https://CRAN.R-project.org/package=NeuralEstimators}}.
%Type = Article
\bibitem[{Sainsbury-Dale et~al.(2025a)Sainsbury-Dale, Richards, Zammit-Mangion
  and Huser}]{Sainsbury-Dale.etal:2024b}
\bibinfo{author}{Sainsbury-Dale, M.}, \bibinfo{author}{Richards, J.},
  \bibinfo{author}{Zammit-Mangion, A.}, \bibinfo{author}{Huser, R.},
  \bibinfo{year}{2025}a.
\newblock \bibinfo{title}{Neural {Bayes} estimators for irregular spatial data
  using graph neural networks}.
\newblock \bibinfo{journal}{Journal of Computational and Graphical Statistics}
  \bibinfo{volume}{34}, \bibinfo{pages}{1153--1168}.
%Type = Unpublished
\bibitem[{Sainsbury-Dale et~al.(2025b)Sainsbury-Dale, Zammit-Mangion, Cressie
  and Huser}]{Sainsbury-Dale_2025_incomplete_data}
\bibinfo{author}{Sainsbury-Dale, M.}, \bibinfo{author}{Zammit-Mangion, A.},
  \bibinfo{author}{Cressie, N.}, \bibinfo{author}{Huser, R.},
  \bibinfo{year}{2025}b.
\newblock \bibinfo{title}{Neural parameter estimation with incomplete data}.
\newblock \bibinfo{note}{ArXiv preprint arXiv.2501.04330}.
%Type = Article
\bibitem[{Sainsbury-Dale et~al.(2024)Sainsbury-Dale, Zammit-Mangion and
  Huser}]{Sainsbury-Dale.etal:2024a}
\bibinfo{author}{Sainsbury-Dale, M.}, \bibinfo{author}{Zammit-Mangion, A.},
  \bibinfo{author}{Huser, R.}, \bibinfo{year}{2024}.
\newblock \bibinfo{title}{Likelihood-free parameter estimation with neural
  {B}ayes estimators}.
\newblock \bibinfo{journal}{The American Statistician} \bibinfo{volume}{78},
  \bibinfo{pages}{1--14}.
%Type = Article
\bibitem[{Scarrott and {MacDonald}(2012)}]{Scarrott.MacDonald:2012}
\bibinfo{author}{Scarrott, C.}, \bibinfo{author}{{MacDonald}, A.},
  \bibinfo{year}{2012}.
\newblock \bibinfo{title}{A review of extreme value threshold estimation and
  uncertainty quantification}.
\newblock \bibinfo{journal}{{REVSTAT}} \bibinfo{volume}{10},
  \bibinfo{pages}{33--60}.
%Type = Article
\bibitem[{Schad et~al.(2021)Schad, Betancourt and Vasishth}]{Schad_2019}
\bibinfo{author}{Schad, D.J.}, \bibinfo{author}{Betancourt, M.},
  \bibinfo{author}{Vasishth, S.}, \bibinfo{year}{2021}.
\newblock \bibinfo{title}{Toward a principled {B}ayesian workflow in cognitive
  science.}
\newblock \bibinfo{journal}{Psychological Methods} \bibinfo{volume}{26},
  \bibinfo{pages}{103}.
%Type = Unpublished
\bibitem[{Shi et~al.(2024)Shi, Zhang, Risser and Shaby}]{shi2024spatial}
\bibinfo{author}{Shi, M.}, \bibinfo{author}{Zhang, L.},
  \bibinfo{author}{Risser, M.D.}, \bibinfo{author}{Shaby, B.A.},
  \bibinfo{year}{2024}.
\newblock \bibinfo{title}{Spatial scale-aware tail dependence modeling for
  high-dimensional spatial extremes}.
\newblock \bibinfo{note}{ArXiv preprint arXiv:2412.07957}.
%Type = Article
\bibitem[{Smith(1990)}]{smith1990extreme}
\bibinfo{author}{Smith, R.L.}, \bibinfo{year}{1990}.
\newblock \bibinfo{title}{Extreme value theory}.
\newblock \bibinfo{journal}{Handbook of Applicable Mathematics}
  \bibinfo{volume}{7}, \bibinfo{pages}{18}.
%Type = Article
\bibitem[{Stein(2021a)}]{stein2021a}
\bibinfo{author}{Stein, M.L.}, \bibinfo{year}{2021}a.
\newblock \bibinfo{title}{A parametric model for distributions with flexible
  behavior in both tails}.
\newblock \bibinfo{journal}{Environmetrics} \bibinfo{volume}{32},
  \bibinfo{pages}{e2658}.
%Type = Article
\bibitem[{Stein(2021b)}]{stein2021b}
\bibinfo{author}{Stein, M.L.}, \bibinfo{year}{2021}b.
\newblock \bibinfo{title}{Parametric models for distributions when interest is
  in extremes with an application to daily temperature}.
\newblock \bibinfo{journal}{Extremes} \bibinfo{volume}{24},
  \bibinfo{pages}{293--323}.
%Type = Misc
\bibitem[{{Swiss Re}(2019)}]{SwissRe:2019}
\bibinfo{author}{{Swiss Re}}, \bibinfo{year}{2019}.
\newblock \bibinfo{title}{Natural catastrophes and man-made disasters in 2018:
  ``secondary'' perils on the frontline}.
\newblock \bibinfo{note}{(Tech. Rep. No. 2)}.
%Type = Article
\bibitem[{Säilynoja et~al.(2022)Säilynoja, B\"urkner and
  Vehtari}]{Sailynoja_2022}
\bibinfo{author}{Säilynoja, T.}, \bibinfo{author}{B\"urkner, P.},
  \bibinfo{author}{Vehtari, A.}, \bibinfo{year}{2022}.
\newblock \bibinfo{title}{Graphical test for discrete uniformity and its
  applications in goodness-of-fit evaluation and multiple sample comparison}.
\newblock \bibinfo{journal}{Statistics and Computing} \bibinfo{volume}{32}.
%Type = Unpublished
\bibitem[{Talts et~al.(2018)Talts, Betancourt, Simpson, Vehtari and
  Gelman}]{Talts_2018}
\bibinfo{author}{Talts, S.}, \bibinfo{author}{Betancourt, M.},
  \bibinfo{author}{Simpson, D.}, \bibinfo{author}{Vehtari, A.},
  \bibinfo{author}{Gelman, A.}, \bibinfo{year}{2018}.
\newblock \bibinfo{title}{Validating {B}ayesian inference algorithms with
  simulation-based calibration}.
\newblock \bibinfo{note}{ArXiv preprint arXiv:1804.06788}.
%Type = Article
\bibitem[{Tencaliec et~al.(2020)Tencaliec, Favre, Naveau, Prieur and
  Nicolet}]{tencaliec2020}
\bibinfo{author}{Tencaliec, P.}, \bibinfo{author}{Favre, A.C.},
  \bibinfo{author}{Naveau, P.}, \bibinfo{author}{Prieur, C.},
  \bibinfo{author}{Nicolet, G.}, \bibinfo{year}{2020}.
\newblock \bibinfo{title}{Flexible semiparametric generalized {Pareto} modeling
  of the entire range of rainfall amount}.
\newblock \bibinfo{journal}{Environmetrics} \bibinfo{volume}{31},
  \bibinfo{pages}{e2582}.
%Type = Inproceedings
\bibitem[{Teshima et~al.(2020)Teshima, Ishikawa, Tojo and
  \emph{et~al.}}]{Teshima_2020}
\bibinfo{author}{Teshima, T.}, \bibinfo{author}{Ishikawa, I.},
  \bibinfo{author}{Tojo, K.}, \bibinfo{author}{\emph{et~al.}},
  \bibinfo{year}{2020}.
\newblock \bibinfo{title}{Coupling-based invertible neural networks are
  universal diffeomorphism approximators}, in: \bibinfo{editor}{Larochelle,
  H.}, \bibinfo{editor}{Ranzato, M.}, \bibinfo{editor}{Hadsell, R.},
  \bibinfo{editor}{Balcan, M.}, \bibinfo{editor}{Lin, H.} (Eds.),
  \bibinfo{booktitle}{Advances in Neural Information Processing Systems}, pp.
  \bibinfo{pages}{3362--3373}.
%Type = Article
\bibitem[{Vicente-Serrano et~al.(2020)Vicente-Serrano, Quiring, Peña-Gallardo,
  Yuan and Domínguez-Castro}]{Vicente-Serrano.etal:2020}
\bibinfo{author}{Vicente-Serrano, S.M.}, \bibinfo{author}{Quiring, S.M.},
  \bibinfo{author}{Peña-Gallardo, M.}, \bibinfo{author}{Yuan, S.},
  \bibinfo{author}{Domínguez-Castro, F.}, \bibinfo{year}{2020}.
\newblock \bibinfo{title}{A review of environmental droughts: Increased risk
  under global warming?}
\newblock \bibinfo{journal}{Earth-Science Reviews} \bibinfo{volume}{201},
  \bibinfo{pages}{102953}.
%Type = Article
\bibitem[{Vrac et~al.(2007)Vrac, Naveau and Drobinsky}]{VracNaveauBrobinsky07}
\bibinfo{author}{Vrac, M.}, \bibinfo{author}{Naveau, P.},
  \bibinfo{author}{Drobinsky, P.}, \bibinfo{year}{2007}.
\newblock \bibinfo{title}{Modeling pairwise rainfall intensities}.
\newblock \bibinfo{journal}{Nonlinear Processes in Geophysics}
  \bibinfo{volume}{14}, \bibinfo{pages}{789--797}.
%Type = Article
\bibitem[{Wadsworth et~al.(2017)Wadsworth, Tawn, Davison and
  Elton}]{wadsworth2017modelling}
\bibinfo{author}{Wadsworth, J.}, \bibinfo{author}{Tawn, J.},
  \bibinfo{author}{Davison, A.}, \bibinfo{author}{Elton, D.},
  \bibinfo{year}{2017}.
\newblock \bibinfo{title}{Modelling across extremal dependence classes}.
\newblock \bibinfo{journal}{Journal of the Royal Statistical Society Series B:
  Statistical Methodology} \bibinfo{volume}{79}, \bibinfo{pages}{149--175}.
%Type = Article
\bibitem[{Wadsworth and Tawn(2022)}]{wadsworth2022higher}
\bibinfo{author}{Wadsworth, J.L.}, \bibinfo{author}{Tawn, J.A.},
  \bibinfo{year}{2022}.
\newblock \bibinfo{title}{Higher-dimensional spatial extremes via single-site
  conditioning}.
\newblock \bibinfo{journal}{Spatial Statistics} \bibinfo{volume}{51},
  \bibinfo{pages}{100677}.
%Type = Article
\bibitem[{Yadav et~al.(2021)Yadav, Huser and Opitz}]{yadav2021spatial}
\bibinfo{author}{Yadav, R.}, \bibinfo{author}{Huser, R.},
  \bibinfo{author}{Opitz, T.}, \bibinfo{year}{2021}.
\newblock \bibinfo{title}{Spatial hierarchical modeling of threshold
  exceedances using rate mixtures}.
\newblock \bibinfo{journal}{Environmetrics} \bibinfo{volume}{32},
  \bibinfo{pages}{e2662}.
%Type = Article
\bibitem[{Yadav et~al.(2022)Yadav, Huser and Opitz}]{yadav2022flexible}
\bibinfo{author}{Yadav, R.}, \bibinfo{author}{Huser, R.},
  \bibinfo{author}{Opitz, T.}, \bibinfo{year}{2022}.
\newblock \bibinfo{title}{A flexible {Bayesian} hierarchical modeling framework
  for spatially dependent peaks-over-threshold data}.
\newblock \bibinfo{journal}{Spatial Statistics} \bibinfo{volume}{51},
  \bibinfo{pages}{100672}.
%Type = Article
\bibitem[{Yadav et~al.(2023)Yadav, Huser, Opitz and Lombardo}]{yadav2023joint}
\bibinfo{author}{Yadav, R.}, \bibinfo{author}{Huser, R.},
  \bibinfo{author}{Opitz, T.}, \bibinfo{author}{Lombardo, L.},
  \bibinfo{year}{2023}.
\newblock \bibinfo{title}{Joint modelling of landslide counts and sizes using
  spatial marked point processes with sub-asymptotic mark distributions}.
\newblock \bibinfo{journal}{Journal of the Royal Statistical Society Series C:
  Applied Statistics} \bibinfo{volume}{72}, \bibinfo{pages}{1139--1161}.
%Type = Incollection
\bibitem[{Yadav et~al.(2025)Yadav, Lombardo and Huser}]{yadav2025statistics}
\bibinfo{author}{Yadav, R.}, \bibinfo{author}{Lombardo, L.},
  \bibinfo{author}{Huser, R.}, \bibinfo{year}{2025}.
\newblock \bibinfo{title}{Statistics of extremes for landslides and
  earthquakes}, in: \bibinfo{editor}{M.~de Carvalho, R.~Huser, P.N.},
  \bibinfo{editor}{Reich, B.} (Eds.), \bibinfo{booktitle}{Handbook on
  Statistics of Extremes}. \bibinfo{publisher}{CRC Press}.
  chapter~\bibinfo{chapter}{27}.
%Type = Inproceedings
\bibitem[{Zaheer et~al.(2017)Zaheer, Kottur, Ravanbakhsh and
  \emph{et~al.}}]{Zaheer_2017_Deep_Sets}
\bibinfo{author}{Zaheer, M.}, \bibinfo{author}{Kottur, S.},
  \bibinfo{author}{Ravanbakhsh, S.}, \bibinfo{author}{\emph{et~al.}},
  \bibinfo{year}{2017}.
\newblock \bibinfo{title}{Deep sets}, in: \bibinfo{editor}{Guyon, I.},
  \bibinfo{editor}{Luxburg, U.V.}, \bibinfo{editor}{Bengio, S.},
  \bibinfo{editor}{Wallach, H.}, \bibinfo{editor}{Fergus, R.},
  \bibinfo{editor}{Vishwanathan, S.}, \bibinfo{editor}{Garnett, R.} (Eds.),
  \bibinfo{booktitle}{Proceedings of the 30th Conference on Neural Information
  Processing Systems}, pp. \bibinfo{pages}{3392--3402}.
%Type = Article
\bibitem[{Zammit-Mangion et~al.(2025)Zammit-Mangion, Sainsbury-Dale and
  Huser}]{Zammit-Mangion.etal:2025}
\bibinfo{author}{Zammit-Mangion, A.}, \bibinfo{author}{Sainsbury-Dale, M.},
  \bibinfo{author}{Huser, R.}, \bibinfo{year}{2025}.
\newblock \bibinfo{title}{Neural methods for amortized inference}.
\newblock \bibinfo{journal}{Annual Reviews of Statistics and Its Application}
  \bibinfo{volume}{12}, \bibinfo{pages}{311--335}.
%Type = Incollection
\bibitem[{Zhang et~al.(2025)Zhang, Rohrbeck and Opitz}]{zhang2025}
\bibinfo{author}{Zhang, L.}, \bibinfo{author}{Rohrbeck, C.},
  \bibinfo{author}{Opitz, T.}, \bibinfo{year}{2025}.
\newblock \bibinfo{title}{Subasymptotic models for spatial extremes}, in:
  \bibinfo{editor}{M.~de Carvalho, R.~Huser, P.N.}, \bibinfo{editor}{Reich, B.}
  (Eds.), \bibinfo{booktitle}{Handbook on Statistics of Extremes}.
  \bibinfo{publisher}{CRC Press}. chapter~\bibinfo{chapter}{17}.

\end{thebibliography}
%\endgroup


%\begin{thebibliography}{00}


\newpage
\appendix
\section{Additional diagnostic plots for the other station pairs}
\label{sec:S-diagnostics}

\subsection*{Ammerzoden--Giersbergen}
\label{sec:S-AG}

\begin{figure}[h!]
  \centering
  % Replace with your actual filename:
  % e.g., full_dashboard_AG.png or .pdf (place in same folder as this .tex)
  \includegraphics[width=\linewidth]{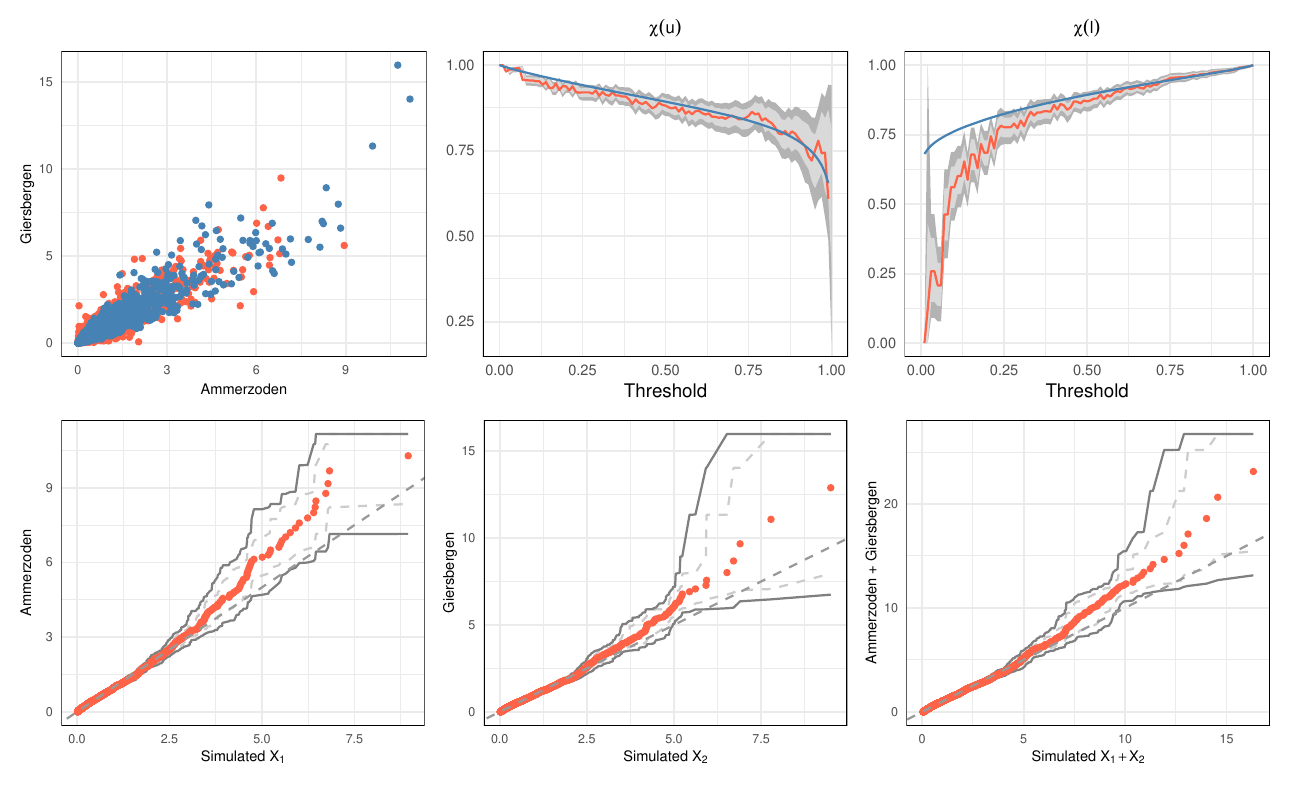}
  \caption{\small Diagnostics for the Ammerzoden--Giersbergen fit.
  Top: Real (red) and simulated (blue) scatterplot based on the estimated parameters (left), and the $\chi$-dependence measures for the upper (middle) and lower (right) tails, computed from observed (red) and simulated (blue) data, with 95\% bootstrap overall error bands (dark gray) and pointwise error bands (light grey).
  Bottom: QQ-plots of observed versus simulated data for each margin (left and middle) and for the joint distribution summarized by the sum across stations (right).
  The 95\% overall error (dark gray line) and the pointwise confidence bands (dashed light gray) were obtained by bootstrapping.
  Recall that the data have been rescaled by their empirical standard deviations.}
  \label{fig:S1_precip_diag_AG}
\end{figure}

\begin{figure}[h!]
  \centering
  % Replace with your actual filename:
  \includegraphics[width=\linewidth]{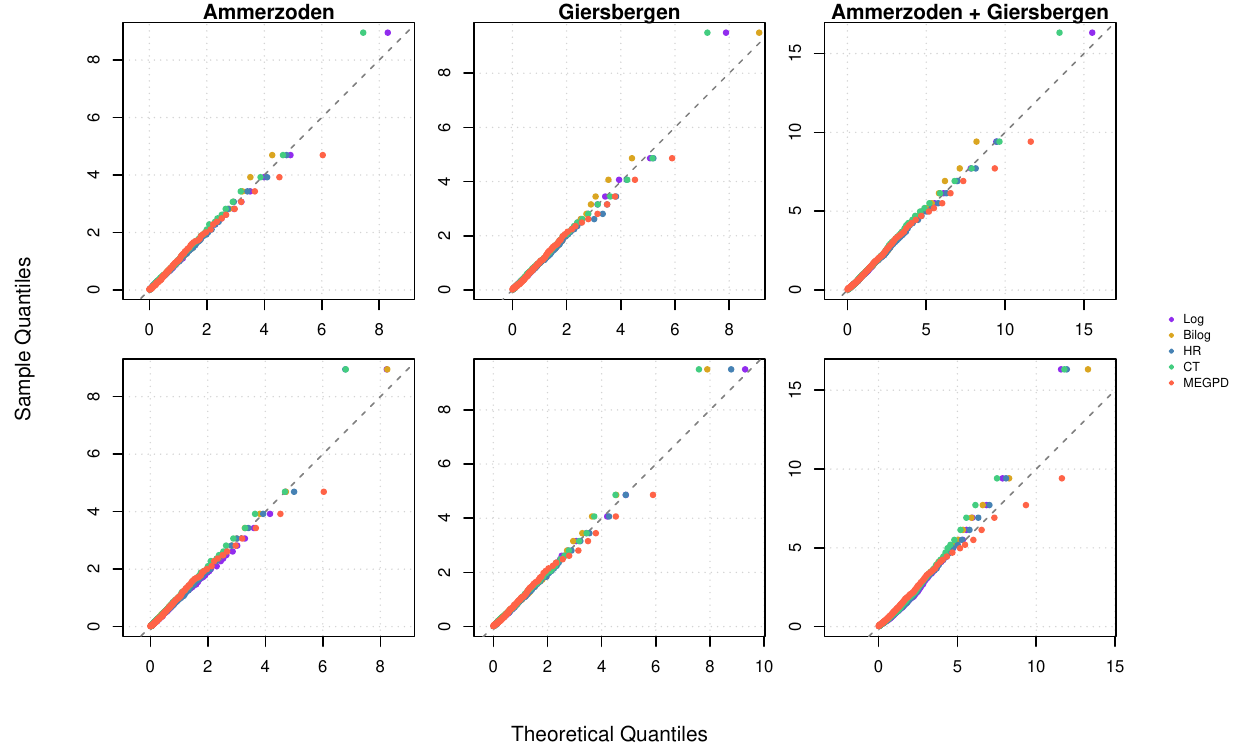}
  \caption{\small QQ-plots for marginal (left and middle) and joint (right; summarized by the sum across stations) distributions, comparing classical bivariate GP models fitted to data from the upper tail (top) or lower tail (bottom), versus our multivariate eGPD model fitted to the entire dataset, for the Ammerzoden--Giersbergen pair. Thresholds are selected as $2.82$ mm (Ammerzoden) and $2.81$ mm (Giersbergen) for the upper tail, and $0.082$ mm (Ammerzoden) and $0.094$ mm (Giersbergen) for the lower tail (after rescaling by the empirical standard deviation). For readability, the QQ-plots are displayed after transforming data from each tail to a common exponential scale (so, for the lower tail, small values are to the right of each panel).
  Classical bivariate GP fits include the logistic (``Log''), bilogistic (``Bilog''), H\"usler--Reiss (``HR''), and Coles--Tawn (``CT'') models, along with our multivariate eGPD model (``MEGPD'').}
  \label{fig:S2_precip_compare_AG}
\end{figure}
%\FloatBarrier

\clearpage
\newpage
\subsection*{Giersbergen--Zaltbommel}
\label{sec:S-GZ}

\begin{figure}[h!]
  \centering
  % Replace with your actual filename:
  \includegraphics[width=\linewidth]{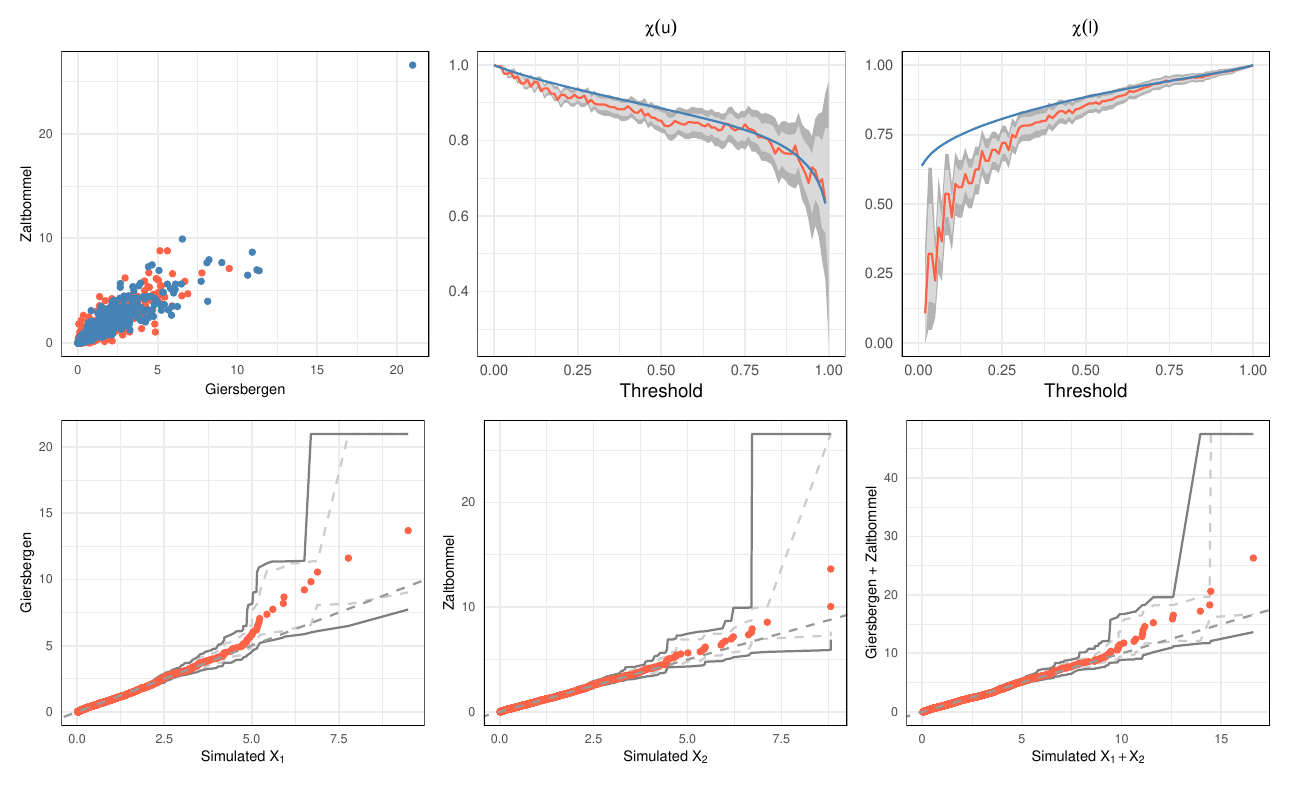}
  \caption{\small Diagnostics for the Giersbergen--Zaltbommel fit.
  Top: Real (red) and simulated (blue) scatterplot based on the estimated parameters (left), and the $\chi$-dependence measures for the upper (middle) and lower (right) tails, computed from observed (red) and simulated (blue) data, with 95\% bootstrap overall error bands (dark gray) and pointwise error bands (light grey).
  Bottom: QQ-plots of observed versus simulated data for each margin (left and middle) and for the joint distribution summarized by the sum across stations (right).
  The 95\% overall error (dark gray line) and the pointwise confidence bands (dashed light gray) were obtained by bootstrapping.
  Recall that the data have been rescaled by their empirical standard deviations.}
  \label{fig:S3_precip_diag_GZ}
\end{figure}
\begin{figure}[h!]
  \centering
  % Replace with your actual filename:
  \includegraphics[width=\linewidth]{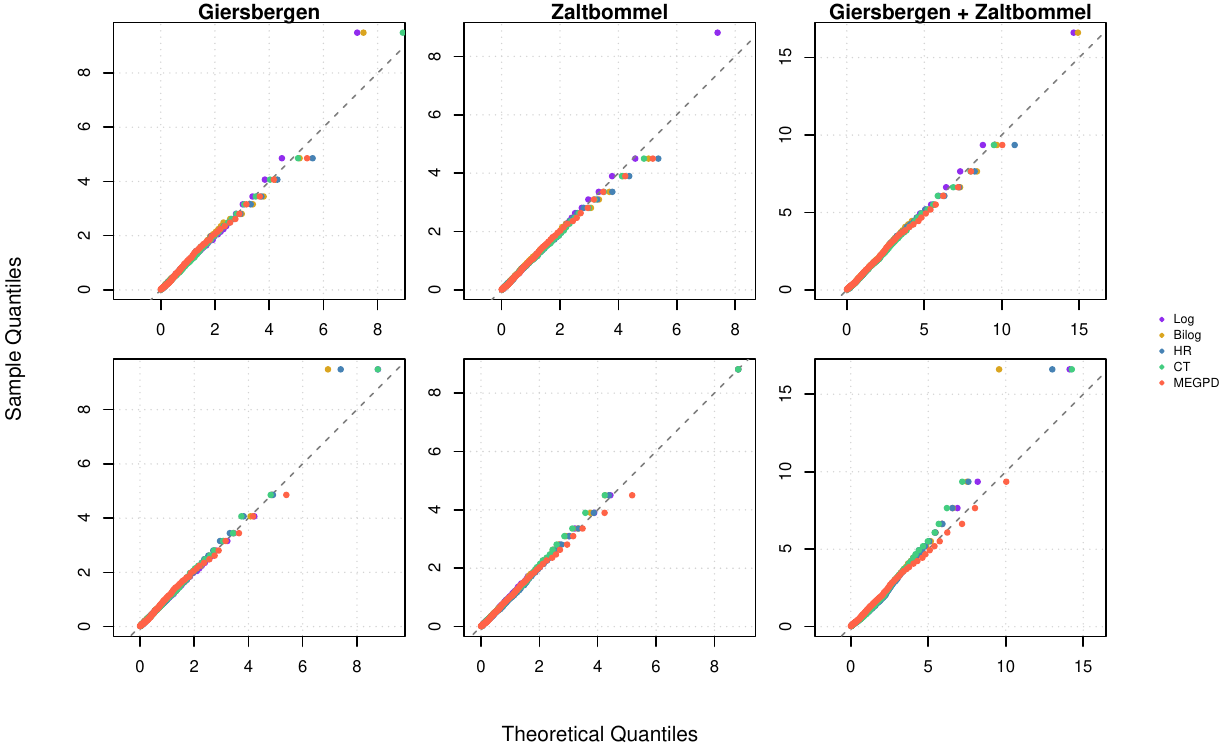}
  \caption{\small QQ-plots for marginal (left and middle) and joint (right; summarized by the sum across stations) distributions, comparing classical bivariate GP models fitted to the upper (top) and lower (bottom) tails versus our multivariate eGPD fitted to the entire dataset, for the Giersbergen--Zaltbommel pair. Thresholds are selected as $2.80$ mm (Giersbergen) and $2.81$ mm (Zaltbommel) for the upper tail (after rescaling by the empirical standard deviation). For readability, the QQ-plots are displayed after transforming data from each tail to a common exponential scale.
  Classical bivariate GP fits include ``Log'', ``Bilog'', ``HR'', and ``CT'', together with ``MEGPD''.}
  \label{fig:S4_precip_compare_GZ}
\end{figure}

%\end{thebibliography}
\end{document}